\documentclass[10.75pt, a4paper]{article}
\usepackage{titlesec}
\titleformat{\section}
  {\bf\sffamily}
  {\thesection. }
  {5pt}
  {\MakeUppercase}
\renewcommand{\thesection}{\Roman{section}} 

\titleformat{\subsection}
  {\bf\sffamily}
  {\thesubsection. }
  {5pt}{}
  
\renewcommand{\thesubsection}{\Alph{subsection}}

\usepackage[affil-it]{authblk} 
\usepackage{etoolbox}
\usepackage{lmodern}

\usepackage{eurosym}

\usepackage[utf8]{inputenc}
\usepackage{geometry}
\usepackage[hidelinks]{hyperref}
\usepackage[citestyle=numeric,style=phys,biblabel=brackets,backend=bibtex,sorting=none,url=false,doi=false, natbib]{biblatex}
\bibliography{bibliography}
\DefineBibliographyStrings{english}{andothers={\itshape et\addabbrvspace al\adddot}}

\usepackage{gensymb} 

\usepackage{csquotes}
\usepackage[]{authblk} 
\usepackage{etoolbox}
\usepackage{lmodern}

\usepackage{amsmath}
\usepackage{enumerate}
\usepackage{enumitem}
\usepackage{graphicx}
\usepackage{siunitx}
\usepackage{float}
\usepackage[]{parskip} 

\makeatletter
\patchcmd{\@maketitle}{\LARGE \@title}{\fontsize{16}{19.2}\selectfont\@title}{}{}
\makeatother

\usepackage[normalem]{ulem}

\usepackage{graphicx}
\usepackage{dcolumn}
\usepackage{bm}
\usepackage{tikz}
\usetikzlibrary{math}
\usetikzlibrary{shapes.geometric, arrows}
\usetikzlibrary{positioning}
\usetikzlibrary{shapes,arrows}
\usetikzlibrary{intersections}
\usepackage{csvsimple}
\usepackage{changepage}
\usepackage[tbtags]{mathtools}
\usepackage{siunitx}

\usepackage[normalem]{ulem}
\usepackage[textwidth=\dimexpr\textwidth-2cm\relax]{todonotes}
\makeatletter
\@mparswitchfalse%
\makeatother
\normalmarginpar 

\usepackage{float}
\usepackage[
	nonumberlist, 				
	acronym,      				
	nomain,						
	nopostdot,					
]{glossaries}
\usepackage{glossary-superragged}
\usepackage[hidelinks]{hyperref}
\usepackage{color, colortbl}

\usepackage{wrapfig}

\usepackage{pgfplots}
\usepackage{tikz}
\usetikzlibrary{arrows}
\usepackage{amsmath}
\pgfplotsset{compat=newest}
\usepgfplotslibrary{fillbetween}
\usetikzlibrary{calc}
\def\centerarc[#1](#2)(#3:#4:#5)
{ \draw[#1] ($(#2)+({#5*cos(#3)},{#5*sin(#3)})$) arc (#3:#4:#5); }

\usepackage{amssymb}

\usepackage{nicefrac}

\usepackage{abstract}

\usepackage{multirow}
\usepackage{tabularx}
\usepackage{booktabs}
\usepackage{array}
\usepackage{longtable}
\newcolumntype{L}[1]{>{\raggedright\let\newline\\\arraybackslash\hspace{0pt}}m{#1}}
\newcolumntype{C}[1]{>{\centering\let\newline\\\arraybackslash\hspace{0pt}}m{#1}}
\newcolumntype{R}[1]{>{\raggedleft\let\newline\\\arraybackslash\hspace{0pt}}m{#1}}

\newacronym{3d}{3D}{three dimensional}
\newacronym{am}{AM}{additive manufacturing}
\newacronym{fdm}{FDM}{fused deposition modeling}
\newacronym{ism}{ISM}{in-space manufacturing}
\newacronym{iss}{ISS}{International Space Station}
\newacronym{fcb}{FCB}{Functional Cargo Block}
\newacronym{dem}{DEM}{discrete element method}
\newacronym{md}{MD}{molecular dynamics}
\newacronym{dc}{DC}{direct-current}
\newacronym[plural=PFCs,firstplural=parabolic flight campaigns (PFCs)]{pfc}{PFC}{Parabolic Flight Campaign}
\newacronym{fft}{FFT}{Fast Fourrier Transform}
\newacronym{cad}{CAD}{Computer Assisted Design}
\newacronym{ptfe}{PTFE}{polytetrafluoroethylene}
\newacronym{ps}{PS}{polystyrene}
\newacronym{nasa}{NASA}{National Aeronautics and Space Administration}
\newacronym{esamm}{ESAMM}{Extended Structure Additive Manufacturing Machine}
\newacronym{amf}{AMF}{Additive Manufacturing Facility}
\newacronym{us}{US}{United States}
\newacronym{usa}{USA}{United States of America}
\newacronym{bmgs}{BMGs}{Bulk Metallic Glasses}
\newacronym{esa}{ESA}{European Space Agency}
\newacronym{si}{SI}{International System of Units, abbreviated from French \textit{Syst\`{e}me International (d'unit\'{e}s)}}
\newacronym{dlr}{DLR}{German Aerospace Center}
\newacronym{liggghts}{LIGGGHTS}{\acrshort{lammps} Improved for General Granular and Granular Heat Transfer Simulations}
\newacronym{lammps}{LAMMPS}{Large-scale Atomic/Molecular Massively Parallel Simulator}
\newacronym{sjkr}{SJKR}{Simplified Johnson-Kendall-Roberts}
\newacronym{ded}{DED}{Directed Energy Deposition}
\newacronym{slm}{SLM}{Selective Laser Melting}
\newacronym{sls}{SLS}{Selective Laser Sintering}
\newacronym{eva}{EVA}{Extra-Vehicular Activity}
\newacronym{sem}{SEM}{Scanning Electron Microscopy}
\newacronym{RPM}{RPM}{Ramdom Positioning Machine}
\newacronym{rpm}{rpm}{revolutions per minute}
\newacronym{rise}{RISE}{Research Internships in Science and Engineering}
\newacronym{daad}{DAAD}{German Academic Exchange Service, abbreviated from German \textit{Deutscher Akademischer Austauschdienst}}
\newacronym{fsm}{FSM}{finite-state machine}
\newacronym{ir}{IR}{infrared}
\newacronym{pcbs}{PCBs}{Printed Circuit Boards}
\newacronym{pcb}{PCB}{Printed Circuit Board}
\newacronym{mcr}{MCR}{Modular Compact Rheometer}
\newacronym{sff}{SFF}{Solid Freeform Fabrication}
\newacronym{uv}{UV}{ultraviolet}
\newacronym{abs}{ABS}{acrylonitrile butadiene styrene}
\newacronym{hpde}{HPDE}{high density polyethylene}
\newacronym{pei}{PEI}{polyetherimide}
\newacronym{bff}{BFF}{BioFabrication Facility}
\newacronym{lens}{LENS}{Laser Engineered Net Shaping}
\newacronym{cnc}{CNC}{Computer Numerical Control}
\newacronym{ebf3}{EBF$^3$}{Electron Beam Free-Form Fabrication}
\newacronym{leo}{LEO}{Low Earth Orbit}
\newacronym{pc}{PC}{polycarbonate}
\newacronym{crissp}{CRISSP}{Customisable Recyclable International Space Station Packaging}
\newacronym{Athena}{Athena}{Advanced Telescope for High-ENergy Astrophysics}
\newacronym{lbm}{LBM}{Laser Beam Melting}
\newacronym{bam}{BAM}{Federal Institute for Materials Research and Testing, abbreviated from German \textit{Bundesanstalt f\"{u}r Materialforschung und-pr\"{u}fung}}
\newacronym{pbf}{PBF}{powder bed fusion}
\newacronym{eb}{EB}{Electron Beam}
\newacronym{2d}{2D}{two dimensional}
\newacronym{4d}{4D}{four dimensional}
\newacronym{ft4}{FT4}{Freeman Technology 4 Powder Rheometer}
\newacronym{dsc}{DSC}{Differential Scanning Calorimetry}
\newacronym{pmma}{PMMA}{polymethylmethacrylate}
\newacronym{1g}{$1g$}{gravity on-ground}
\newacronym{mug}{$\mu g$}{microgravity}
\newacronym{bcm}{BCM}{Box Counting Method}
\newacronym{mct}{MCT}{Mode Coupling Theory}
\newacronym{gmct}{gMCT}{granular Mode Coupling Theory}
\newacronym{itt}{ITT}{Integration Through Transients}
\newacronym{mfc}{MFC}{Mass Flow Controller}
\newacronym{ct}{CT}{computed tomography}
\newacronym{xct}{XCT}{X-ray computed tomography}
\newacronym{cv}{CV}{curriculum vitae}
\newacronym{pi}{PI}{principal investigator}
\newacronym{osp}{OSP}{orthogonal superimposed perturbation}
\newacronym{npi}{NPI}{Network Partnering Initiative}
\newacronym{ecsat}{ECSAT}{European Centre for Space Applications and Telecommunications}
\newacronym{eac}{EAC}{European Astronaut Centre}
\newacronym{estec}{ESTEC}{European Space Research and Technology Centre}
\newacronym{fps}{fps}{frames per second}
\newacronym{pdf}{pdf}{probability density function}
\newacronym{al}{Al}{aluminium}
\newacronym{ss}{\textit{SS}}{\textit{Smooth Surface}}
\newacronym{rs}{\textit{RS}}{\textit{Rough Surface}}
\newacronym{rcp}{rcp}{random close packing}
\newacronym{iop}{IoP UvA}{Institute of Physics of the University of Amsterdam}
\newacronym{mp}{MP}{Institute of Material Physics for Space}
\newacronym{elgra}{ELGRA}{European Low Gravity Research Association}
\newacronym{zarm}{ZARM}{Center of Applied Space Technology and Microgravity}
\newacronym{piv}{PIV}{particle image velocimetry}
\usepackage[framemethod=tikz]{mdframed}
\usepackage{lipsum}
\usepackage{enumitem}
\usepackage{mdframed}
\usepackage{amsmath}
\usepackage{physics}
\usepackage{setspace}
\usepackage{mathtools}
\usepackage{hyperref}
\usepackage{graphicx}
\usepackage{caption}
\usepackage{subcaption}
\usepackage{wasysym}
\usepackage{bm}
\usepackage{xcolor}
\usepackage[english]{babel}
\usepackage[nottoc]{tocbibind}
\usepackage{amssymb}

\usepackage[most]{tcolorbox}
\newtcolorbox{myBox}[3][]{
arc=2mm,
lower separated=true,
fonttitle=\bfseries,
colbacktitle=gray!10,
coltitle=black!50!black,
enhanced,
colframe=gray!10,
colback=gray!10,
title=#2,#1}

\usepackage{dirtytalk}
\usepackage{tcolorbox}
\usepackage[font=footnotesize]{caption}
\newtcolorbox{mybox}[1]{colback=green!6!white,colframe=black!75!black,fonttitle=\bfseries,title=#1}
\newtcolorbox{mybox2}{colback=red!5!white,colframe=red!75!black}

\usepackage{pifont}

\usepackage{soul,xcolor}
\setstcolor{red}

\usepackage{xcolor,hyperref}
\hypersetup{
   colorlinks,
   linkcolor={blue!50!black},
   citecolor={blue!50!black},
   urlcolor={blue!80!black}
} 

\definecolor{mycolor}{rgb}{0.122, 0.435, 0.698}

\usepackage{textgreek}

\title{Droplet Deformation and Emulsion Rheology\\
in Two-Dimensional Odd Stokes Flow}
\author[1]{Thomas Appleford 
\footnote{t.appleford@uva.nl}
}

\affil[1]{Van der Waals-Zeeman Institute, Institute of Physics, University of Amsterdam, The Netherlands}

\author[1]{Hugo Franca
\footnote{h.franca@uva.nl}
}

\author[1]{Maziyar Jalaal 
\footnote{m.jalaal@uva.nl, ORCID: 0000-0002-5654-8505}
}
\begin{document}
    \definecolor{brickred}{rgb}{0.8, 0.25, 0.33}
\definecolor{darkorange}{rgb}{1.0, 0.55, 0.0}
\definecolor{persiangreen}{rgb}{0.0, 0.65, 0.58}
\definecolor{persianindigo}{rgb}{0.2, 0.07, 0.48}
\definecolor{cadet}{rgb}{0.33, 0.41, 0.47}
\definecolor{turquoisegreen}{rgb}{0.63, 0.84, 0.71}
\definecolor{sandybrown}{rgb}{0.96, 0.64, 0.38}
\definecolor{blueblue}{rgb}{0.0, 0.2, 0.6}
\definecolor{ballblue}{rgb}{0.13, 0.67, 0.8}
\definecolor{greengreen}{rgb}{0.0, 0.5, 0.0}
    \begingroup
    \sffamily
    \date{}
    \maketitle
    \endgroup
    
    \begin{abstract}
        We study the deformation of a two-dimensional viscous droplet in simple shear in the presence of odd viscosity. 
        We derive an analytical solution for the droplet shape and surrounding flow field within the framework of odd Stokes flow, allowing for differences in both even and odd viscosity between the droplet and the surrounding fluid.
        This solution yields closed-form expressions for the macroscopic apparent even and odd viscosities of a dilute emulsion.
        We show that, provided all viscosity differences remain moderate, the steady-state Taylor deformation parameter satisfies
        $D_T^\infty = \text{Ca} + \mathcal{O}(\text{Ca}^2)$
        so that the leading-order droplet deformation is unchanged from the classical (even-viscous) result. 
        Nevertheless, pronounced effects emerges beyond leading order, where our direct numerical simulations reveal odd-viscous differences to the droplet deformation.
        In addition, we show that the flow is influenced only by the difference in odd viscosity between the droplet and the medium and not on their individual values.
        Our analysis clarifies how odd viscosity might modify the effective rheology of dilute emulsions and provides a framework for interpreting droplet-based measurements of odd-viscous response.
        
    \end{abstract}  

\section{Introduction}

    In standard fluid mechanics, viscosity is often treated as a scalar quantity, but more generally, it may be considered a tensorial quantity with symmetric and antisymmetric parts. 
    The Onsager relations force the antisymmetric part to vanish, leaving only the dissipative symmetric component. However, Avron~\cite{Avron1998} showed that in two dimensions, if a fluid breaks time-reversal symmetry, the antisymmetric “odd” part need not vanish, so the viscosity tensor acquires non-dissipative components. These components are known as odd viscosity, and are common in chiral fluids.
    Banerjee \emph{et al}.~\cite{Banerjee2017} showed how odd viscosity arises naturally in systems of self-spinning objects such as chiral grains or colloidal particles subject to torques, and odd viscosity has since been observed experimentally in a quasi-two-dimensional fluid composed of spinning colloidal particles confined to a surface (a chiral spinner fluid)~\cite{Soni2019}. 
    A comprehensive review of odd viscosity (and elasticity) and their applications is provided in Fruchart \emph{et al}.~\cite{Fruchart} including the theoretical low-Reynolds-number hydrodynamics that forms the basis of the present work. 
    Of particular relevance here, is the work of Khain \emph{et al}.~\cite{Khain_Scheibner_Fruchart_Vitelli_2022} who developed Green’s functions for odd Stokes flows and found analytical solutions for both the flow past a sphere and a bubble, as well as conducting a numerical study of many-particle sedimentation. 
    A recurring theme in these low-Reynolds-number studies is the generation of lift on immersed bodies due to odd viscosity: Kogan~\cite{Kogan} used the Osseen equations to compute the lift on a rigid cylinder, Hosaka \emph{et al}.~\cite{hosaka2021hydrodynamic} used the linearised Navier–Stokes equations to compute the lift on a fluid cylinder and more recent works by Lier have extended this to a sphere~\cite{Lier_2023}, including finite-Reynolds-number corrections~\cite{Lier_2024}. 
    Odd viscosity has also been shown to influence the locomotion of microswimmers~\cite{Lapa,hosaka2024chirotactic}.

    Aside from being realised in experiments~\cite{Soni2019} and their potential applications in controlling flows, two-dimensional (2D) odd viscous flows are also of inherent theoretical interest. 
    Unlike in higher dimensions, 2D odd viscous fluids are compatible with isotropy~\cite{Avron1998}. 
    Furthermore, the stress tensor may be written as the sum of a null (divergence-free) part, which does not contribute to momentum balance, and an isotropic part that can be accommodated as a modified pressure term~\cite{Kirkinis}. 
    Consequently, flows specified by velocity boundary conditions only are unaffected by odd viscosity.
    On the other hand, flows with stress boundary conditions are influenced by odd viscosity.
    This makes, for example, two-phase flows in 2D a particularly important setting for observing the effects of odd viscosity on the flow~\cite{francca2025odd}.

    With the problem of uniform flow past a fluid cylinder having been solved~\cite{hosaka2021hydrodynamic}, it is natural to next address the problem of a cylindrical droplet in a higher-order flow. 
    In this work, we consider a droplet in simple shear. 
    As a foundational problem in emulsion rheology, the droplet in simple shear has received considerable attention in the literature. 
    The subject has been reviewed many times, notably in the articles by Acrivos~\cite{acrivos1983breakup}, Rallison~\cite{rallison1984deformation}, and Stone~\cite{stone1994dynamics}. 
    Most relevant to the present work are the seminal studies of Taylor~\cite{Taylor1932, Taylor1934}, which use Lamb’s general solution of the Stokes equations~\cite{Lamb} to obtain the flow around a 3D droplet in a purely extensional flow. 
    The degree of deformation was quantified using the eponymous Taylor deformation parameter $D_T = (l - b)/(l + b)$, where $l$ and $b$ denote the longest and shortest droplet dimensions, respectively.
    It was found that provided the viscosity ratio $\lambda \sim 1$ and $\text{Ca} \ll 1$, the steady state value $D_T^\infty$ of $D_T$ was $D_T^\infty = g^{3D}(\lambda) \text{Ca}$ where $\text{Ca}$ is the capillary number and $g^{3D}(\lambda) = (19\lambda + 16)/(16\lambda + 16)$.
    Furthermore, the apparent viscosity $\mu_*$ of a dilute emulsions was found to be $\mu_* = \mu_M(1 + f^{3D}(\lambda) \phi)$ where $\mu_M$ is the matrix viscosity, $\phi$ is the coverage fraction of the droplets and $f^{3D}(\lambda) = (5\lambda + 2)/(2\lambda + 2)$. 
    Our recent work~\cite{appleford2025rheology} presents the 2D version of Taylor's theory and finds $f^{2D}(\lambda) = (2\lambda + 1)(\lambda + 1)$ and $g^{2D} = 1$
    (see also~\cite{richardson1968, richardson1973two, buckmaster1973bursting} for earlier analytical treatments involving complex variables).
    Notably, in the 2D case, the small deformation theory lacks a $\lambda$-dependence. 
    
    The primary objective of the present study is to determine the effect of odd viscosity on the droplet in simple shear system.
    To this end, we first introduce the problem in Section II.\ref{sec:prob}. 
    In Section~II.\ref{sec:eom}, we introduce the equations of motion for a 2D odd fluid, as well as the general solution of the odd Stokes equations.
    In Section~III.\ref{sec:weak_shear}, we solve the droplet-in-shear problem analytically in the limit of weak shear. 
    In Section~III.\ref{sec:moderate_shear}, we employ direct numerical simulations to probe higher shear rates and larger droplet deformations, which are inaccessible via analytical methods.
    In Section~\ref{sec:apparent_viscosity}, we obtain a formula for the apparent viscosity $\mu_*$ of a dilute emulsion.
    Furthermore, we introduce the notion of the apparent odd viscosity $\mu_*^O$ of an emulsion, for which we find a similar expression.
        
\section{The Droplet in Shear}

\subsection{Problem Statement}
\label{sec:prob}

    \begin{figure}[h]
        \centering
        \includegraphics[width=0.8\linewidth]{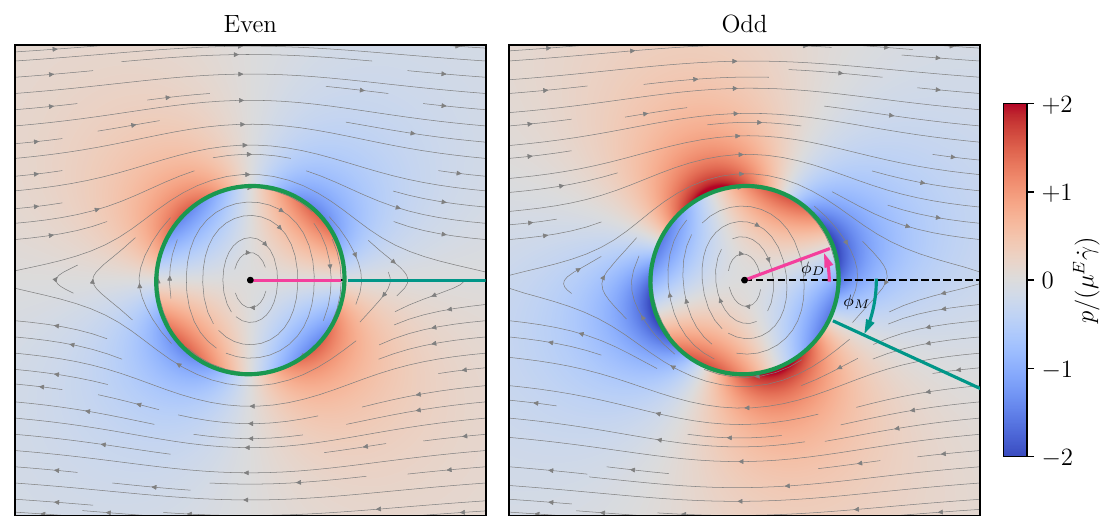}

        \caption{
            The pressure field of two droplets under shear in steady state. 
            Here the shear flow is weak $\text{Ca} \ll 1$ and the droplet (green solid line) remains circular.
            Left: both phases are even.
            Right: both phases are odd.
            Odd viscosity modulates the intensity of pressure fields and causes phase shifts $\phi_D$ and $\phi_M$ in the droplet and the surrounding matrix pressure fields, respectively.}
        \label{fig:intro}
    \end{figure}
    Here we briefly introduce the droplet in shear problem. 
    A more comprehensive introduction, including its implementation in numerical simulations may be found in Appendix~\ref{appendix:numerics}.
    We consider a single circular droplet of radius $R$ placed at the centre of an fluid domain $\Omega$.
    The domain is divided into two disjoint regions: the droplet, $\Omega_D$, and the surrounding matrix, $\Omega_M$.
    In general, the droplet and matrix may have different material properties.
    We denote the mass density, even viscosity and odd viscosity of droplet and matrix by $\rho_D, \mu_D^E, \mu_D^O$ and $\rho_M, \mu_M^E, \mu_M^O$ respectively.
    We denote the surface tension at the interface by $\sigma$.
    The system is controlled by the Reynolds number $\text{Re} = \rho_M \dot{\gamma} R^2/\mu_M^E$, the capillary number $\text{Ca} = \mu_M^E \dot{\gamma} R/\sigma$, the even viscosity ratio $\lambda = \mu_D^E/\mu_M^E$, density ratio $\alpha = \rho_D/\rho_M$ and the relative oddness parameters $\beta_D = \mu_D^O/\mu_D^E$ and $\beta_M = \mu_M^O/\mu_M^E$.
    In our theory, as is standard, we assume $\text{Re} = 0$ and $\text{Ca} \ll 1$, corresponding to a small droplet in a highly viscous liquid subject to weak shear.
    For simplicity, we also take $\lambda = 1$ and $\alpha = 1$ such that each phase has equal even viscosity and density and assume $\Omega $ to be unbounded ($\Omega = \mathbb{R}^2$).

    We focus primarily on the system in steady state, which is reached when the droplet deforms by an amount such that the change in capillary stress matches the viscous stress jump across the interface (see Sec.~\ref{appendix:time-dependence} for a discussion of the transient dynamics).
    Provided the shear is weak enough ($\text{Ca} \ll 1$), the steady-state droplet shape is well approximated by an ellipse.
    Our primary aim is to determine how these additional ``odd'' parameters influence the steady state of the droplet-under-shear system.
    The primary effects of odd viscosity are to phase-shift and rescale the stress fields in the fluid, as depicted in Fig.~\ref{fig:intro} for weak shear $\text{Ca} \ll 1$.
    
\subsection{The Equations of Motion}
    \label{sec:eom}
    The equations of motion for an odd fluid undergoing incompressible flow are the continuity equation and the odd Navier-Stokes equation.
    In two dimensions, these read
    \begin{align}
        \bm{\nabla} \cdot \bm{u} &= 0,
        \label{eqn:continuity}
        \\
        \rho \frac{D \bm{u}}{D t} &= - \bm{\nabla} p
        + \mu^E \nabla^2 \bm{u}
        + \mu^O \bm{\nabla} \omega,
        \label{eqn:odd_navier_stokes}
    \end{align}
    where $\rho$ is the mass density, $D/Dt = \partial/\partial t + \bm{u} \cdot \bm{\nabla}$ is the material derivative, $\mu^E$ is the even viscosity, $\mu^O$ is the odd viscosity, and $\omega = \bm{\epsilon} : \bm{\nabla u}$ is the scalar vorticity, with $\bm{\epsilon}$ the two-dimensional Levi-Civita symbol.
    A derivation of this particular form of the momentum equation may be found in \cite{Fruchart}.
    Since the primary concern of the present work is a system in steady state at zero Reynolds number, it suffices to consider the steady odd Stokes equation, which reads
    \begin{align}
        \bm{\nabla} p &= \mu^E \nabla^2 \bm{u} + \mu^O \bm{\nabla} \omega.
        \label{eqn:odd_stokes}
    \end{align}
    This equation is linear, and its Green's functions are known in both two and three dimensions \cite{Khain_Scheibner_Fruchart_Vitelli_2022}.
    Solutions to problems involving odd creeping flow are therefore often readily attainable via boundary integral methods \cite{Yuan} and may be cast as singularity solutions.
    In this work, however, our primary methods of solution are twofold.
    The first is a modified form of Lamb's solution \cite{appleford2025rheology, Lamb}, adapted to two-dimensional odd flows.
    In the polar coordinate system, $\bm{r} = (r, \theta)$, with $\bm{u} = u_r \bm{\hat{r}} + u_\theta \bm{\hat{\theta}}$, the modified Lamb solution reads
    \begin{align}
        u_r - u_r^\infty
        &=
        \sum_{n=-\infty}^{\infty}
        \left\{
            \frac{n \,\phi^n}{r^2}
            +
            \frac{1}{4 \mu^E}
            \frac{n \, p^n}{n+1}
        \right\}
        \label{eqn:lamb_u_r},
        \\
        u_\theta - u_\theta^\infty
        &=
        \sum_{n=-\infty}^{\infty}
        \left\{
            \frac{1}{r^2} 
            \pdv{\phi^n}{\theta}
            +
            \frac{n+2}{4 \mu^E n (n+1)}
            \pdv{p^n}{\theta}
        \right\}
        \label{eqn:lamb_u_t},
        \\
        p - p^\infty
        &=
        \sum_{n=-\infty}^{\infty}
        \left\{
            p^n
            +
            \frac{\mu^O}{\mu^E}
            \frac{1}{n}
            \pdv{p^n}{\theta}
        \right\},
        \label{eqn:lamb_p}
    \end{align}
    where $p^n$ and $\phi^n$ are circular harmonic functions, which in general may be written as linear combinations of $r^n \sin(n\theta)$ and $r^n \cos(n\theta)$.
    Given the geometry of a problem, obtaining the solution for creeping flow is thus reduced to a careful choice of $p^n$ and $\phi^n$ to match the desired boundary conditions.
    Clearly, this is simplest when the geometry has circular symmetry, but in near-circular geometries such as the droplet under shear, it is possible to construct perturbative solutions, as we will soon demonstrate.
    For large deformations, the perturbative method becomes cumbersome, so we employ a second method of solution, namely direct numerical simulations (DNS) of Eqs.~\eqref{eqn:continuity} and~\eqref{eqn:odd_navier_stokes}.
    We choose a small value of the Reynolds number, $\text{Re} = 0.01$, such that the solutions approximate creeping flow, and a finite domain of side length $L/R = 20 \gg 1$ such that the droplet does not interact with the boundaries.
    A full discussion of the implementation and validation of the simulations may be found in Appendix~\ref{appendix:numerics}.

\clearpage
\section{Droplets in Simple Shear: Steady State Solutions}

\subsection{Weak Shear Flows: Analytical Solution} 
    \label{sec:weak_shear}
    \begin{figure}[h]
        \centering
        \begin{subfigure}{\textwidth}
            \includegraphics[width=\linewidth]{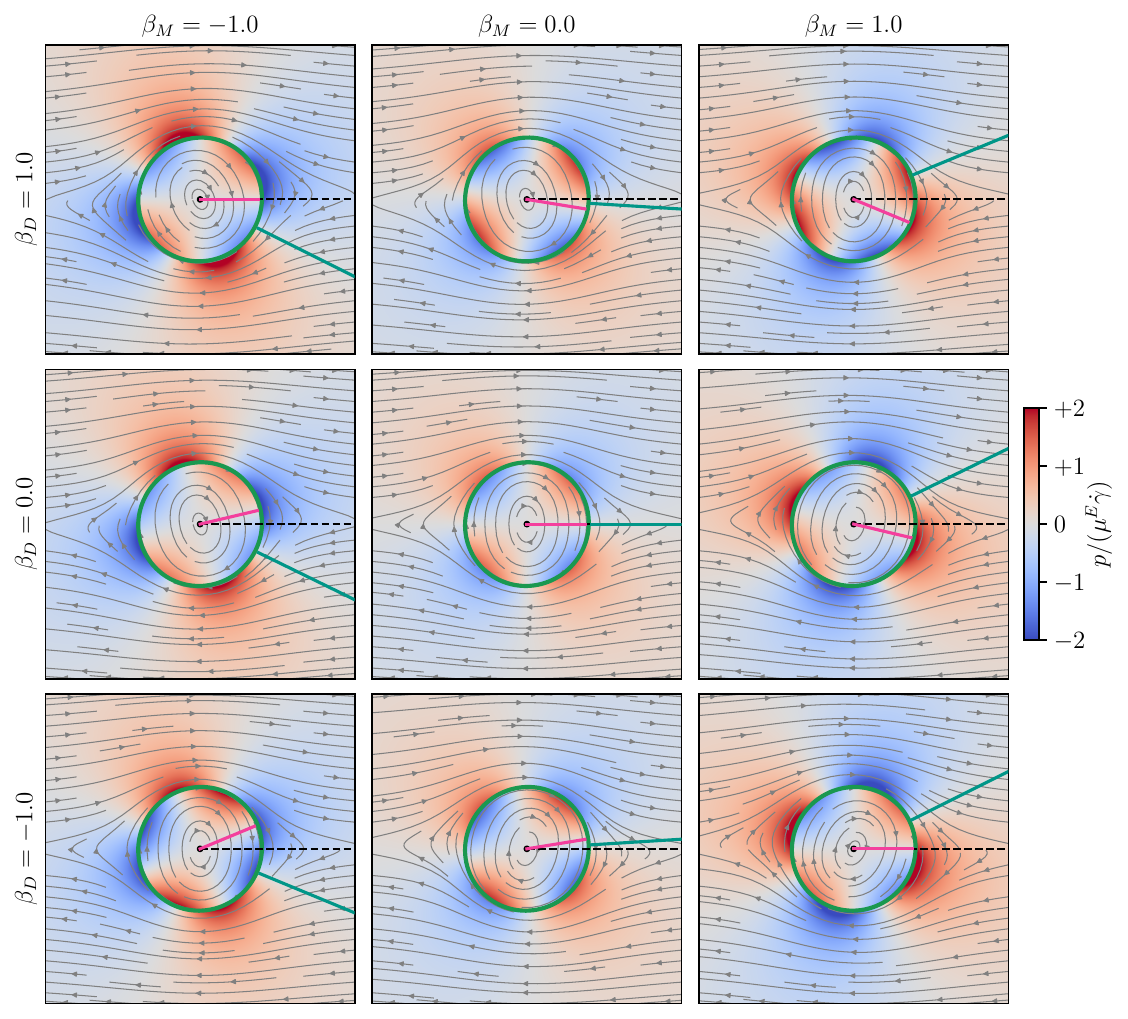}
            \caption{}
        \end{subfigure}
        \begin{subfigure}{0.475\textwidth}
            \includegraphics[width=\linewidth]{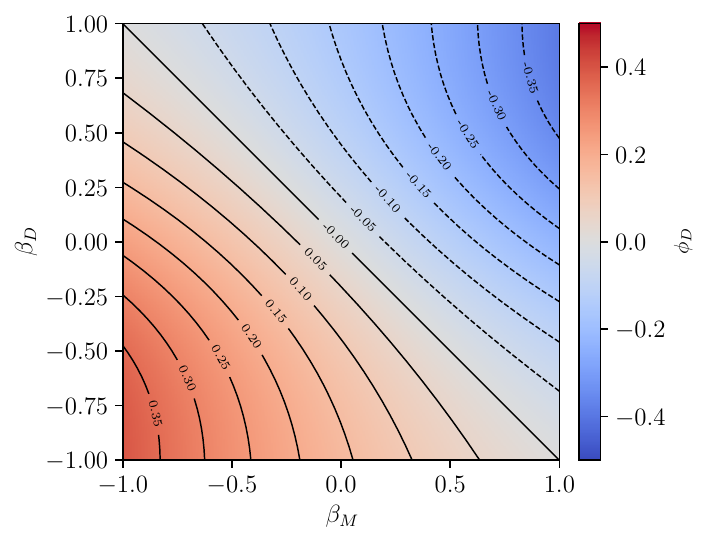}
            \caption{}
        \end{subfigure}
        \begin{subfigure}{0.475\textwidth}
            \includegraphics[width=\linewidth]{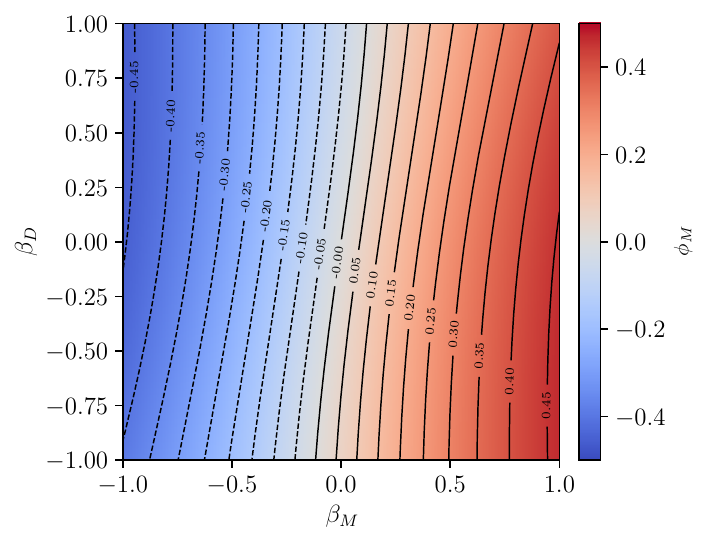}
            \caption{}
        \end{subfigure}

        \caption{
            (a) The pressure field $p$ for an odd-viscous droplet in an odd-viscous simple shear flow according to the analytical solution.
            The Laplace pressure, $\sigma/R$, has been subtracted from the droplet pressure field.
            The parameters are $\text{Ca} = 0.01$, $\lambda = 1$ and $L/R = 20$.
            Panels show different combinations of the oddness parameters $(\beta_D, \beta_M) \in [-1, 1] \times [-1, 1]$.
            The plotted region is $[-2.5R, 2.5R] \times [-2.5R, 2.5R]$.
            The grey curves denote streamlines of the velocity field $\bm{u}$.
            (b) The rotation angle $\phi_D$ (pink line) of the droplet pressure relative to the Newtonian case for oddness parameters $(\beta_D, \beta_M) \in [-1, 1] \times [-1, 1]$. 
            (c) The rotation angle $\phi_M$ (blue-green line) of the matrix pressure relative to the Newtonian case for oddness parameters $(\beta_D, \beta_M) \in [-1, 1] \times [-1, 1]$.
        }
        \label{fig:pressure_Ca=0.01}
    \end{figure}

    We derive leading-order analytical expressions for the shape of the droplet, as well as the flow fields inside and outside the droplet.
    Adopting the standard 2D polar coordinate system, $\bm{r} = (r, \theta)$, we assume the undisturbed flow to be of the form:
    \begin{align}
        u_r^\infty &= \frac{\dot{\gamma}r}{2} \sin(2\theta)  \\
        u_\theta^\infty &= \frac{\dot{\gamma}r}{2} \cos(2\theta)  - \frac{\dot{\gamma}r}{2}
    \end{align}
    with constant pressure, $p^\infty$. 
    This is a simple shear flow with shear rate $\dot{\gamma}$. 
    We assume the droplet interface to be the closed contour $r = \xi(\theta)$ which is centred on the origin $r = 0$.
    At the interface, we impose the continuity of the velocity field and the stress field as well as the kinematic boundary condition.
    We assume that the shape of the droplet is determined by a balance between surface tension and viscous forces.
    This requires that the imposed shear be sufficiently weak that the viscous stresses tending to deform the droplet are counterbalanced by only a small perturbation of its shape.
    Formally, we consider the regime $\text{Ca} \ll 1$ with $\lambda \ll 1/\text{Ca}$ and with $\beta_D$ and $\beta_M$ both finite.
    We note that it is also possible for the droplet to reach a steady state in the absence of surface tension ($\text{Ca} \to \infty$), provided the $\lambda$ is sufficiently large \cite{appleford2025rheology, rallison1984deformation}, but we will not consider that regime here.
    When $\text{Ca} \ll 1$, it is reasonable to assume that the droplet shape changes by some small amount proportional to $\text{Ca}$.
    Motivated by the form of the undisturbed flow, we suggest:
    \begin{align}
        \xi(\theta) &= R\left(1 + \text{Ca} A_2 \sin(2\theta) + \text{Ca} B_2\cos(2\theta)\right) + \mathcal{O}(\text{Ca}^2).
        \label{eqn:ansatz_shape}
    \end{align}
    where $A_2$ and $B_2$ are dimensionless quantities to be determined by the boundary conditions.
    Note that $R$ denotes the equivalent radius of the droplet, defined such that the area of the droplet $\pi R^2 + \mathcal{O}(\text{Ca}^2)$.
    In 2D purely even-viscous simple shear, $A_2 = 1$ and $B_2 = 0$ indicating that, to first order in $\text{Ca}$, the droplet is elliptical with its long axis parallel to the principal eigenvector of the undisturbed shear flow \cite{appleford2025rheology}.
    Note that this does not depend on the viscosity ratio $\lambda$.
    We include $B_2$ to allow for the possibility that the orientation angle and degree of deformation change due to odd viscosity.
    Recall that the capillary stress associated with $\xi(\theta)$ is $\sigma \kappa(\theta)$ where $\sigma$ is the coefficient of surface tension and $\kappa(\theta)$ is the mean curvature of $\xi(\theta)$.
    It is easily shown that:
    \begin{align}
        \sigma \kappa(\theta) 
        &= 
        \frac{\sigma}{R}
        \left(
            1 
            -
            3 \text{Ca} A_2\sin(2\theta) 
            -
            3 \text{Ca} B_2\cos(2\theta)
        \right)
        +
        \mathcal{O}(\text{Ca}^2)
        \label{eqn:capillary_stress}
    \end{align}
    Now we propose an ansatz for the flow fields.
    In shear flow, it is natural to choose harmonics of the following form:
    \begin{align}
        p^n
        &=
        \mu^E \dot{\gamma}
        \left( \frac{r}{R} \right)^n
        \left[
            a_n \sin(n\theta)
            +
            c_n \cos(n\theta)
        \right]
        \label{eqn:ansatz_harmonics_p}
        \\
        \phi^n
        &=
        \dot{\gamma} R^2
        \left( \frac{r}{R} \right)^n
        \left[
            b_n \sin(n\theta)
            +
            d_n \cos(n\theta)
        \right]
        \label{eqn:ansatz_harmonics_phi}
    \end{align}
    where $a_n, b_n, c_n$ and $d_n$ are dimensionless constants to be determined by boundary conditions, and the factors of $\mu^E$ and $R$ are chosen to give $p^n$ and $\phi^n$ the correct dimensions.
    The structure of the undisturbed flow suggests we choose $n = \pm2$.
    For the flow in $\Omega_D$ we choose $n = +2$ such that the flow is regular at the origin whilst in $\Omega_M$ we choose $n = -2$ so that the flow vanishes at infinity.
    Substituting Eqs.~\eqref{eqn:ansatz_harmonics_p} and~\eqref{eqn:ansatz_harmonics_phi} into Eqs.~\eqref{eqn:lamb_u_r},~\eqref{eqn:lamb_u_t} and~\eqref{eqn:lamb_p} and keeping terms up to $\mathcal{O}(\text{Ca})$, we find for the droplet:
    \begin{align}
        u_r^D
        -
        u_r^\infty
        &=
        \frac{\dot{\gamma} r}{2}
        \left[
            4b_2
            +
            \frac{a_2}{3}
            \left( \frac{r}{R} \right)^2
        \right]
        \sin(2\theta)
        +
        \frac{\dot{\gamma} r}{2}
        \left[
            4d_2
            +
            \frac{c_2}{3}
            \left( \frac{r}{R} \right)^2
        \right]
        \cos(2\theta)
        \label{eqn:u_r_int}
        \\
        u_\theta^D
        -
        u_\theta^\infty
        &=
        \frac{\dot{\gamma} r}{2}
        \left[
            4b_2
            +
            \frac{2a_2}{3}
            \left( \frac{r}{R} \right)^2
        \right]
        \cos(2\theta)
        -
        \frac{\dot{\gamma} r}{2}
        \left[
            4d_2
            +
            \frac{2c_2}{3}
            \left( \frac{r}{R} \right)^2
        \right]
        \sin(2\theta)
        \label{eqn:u_t_int}
        \\
        p^D
        -
        p^\infty
        &=
        \frac{\sigma}{R}
        +
        \dot{\gamma}
        \left(\frac{r}{R}\right)^2 
        \left[
            (a_2 \mu^E_D - c_2 \mu^O_D) \sin(2\theta)
            +
            (c_2 \mu^E_D + a_2 \mu^O_D ) \cos(2\theta)
        \right]
        \label{eqn:p_int}
    \end{align}
    and for the matrix:
    \begin{align}
        u_r^M
        -
        u_r^\infty
        &=
        -
        \frac{\dot{\gamma} r}{2}
        \left[
            a_{-2}
            \left( \frac{R}{r} \right)^2
            -
            4b_{-2}
            \left( \frac{R}{r} \right)^4
        \right]
        \sin(2\theta)
        +        
        \frac{\dot{\gamma} r}{2}
        \left[
            c_{-2}
            \left( \frac{R}{r} \right)^2
            -
            4d_{-2}
            \left( \frac{R}{r} \right)^4
        \right]
        \cos(2\theta)
        \label{eqn:u_r_ext}
        \\
        u_\theta^M
        -
        u_\theta^\infty
        &=
        -
        \frac{\dot{\gamma} r}{2}
        \left[
            4b_{-2}
            \left( \frac{R}{r} \right)^4
        \right]
        \cos(2\theta)
        -
        \frac{\dot{\gamma} r}{2}
        \left[
            4d_{-2}
            \left( \frac{R}{r} \right)^4
        \right]
        \sin(2\theta)
        \label{eqn:u_t_ext}
        \\
        p^M
        -
        p^\infty
        &=
        \dot{\gamma}
        \left(\frac{R}{r}\right)^2 
        \left[
            \left(
                -
                a_{-2} \mu^E_M 
                +
                c_{-2} \mu^O_M
            \right)
            \sin(2\theta)
            +
            \left(
                c_{-2} \mu^E_M 
                +
                a_{-2} \mu^O_M
            \right)
            \cos(2\theta)
        \right]
        \label{eqn:p_ext}
    \end{align}
    Note that the Laplace pressure term $\sigma/R$ appearing in Eq.~\eqref{eqn:p_int} arises naturally as the solution to the droplet-under-shear problem in the limiting case $\text{Ca} = 0$ where the shear rate vanishes.
    Imposing the continuity of velocity, stress, as well as the kinematic boundary condition up to $\mathcal{O}(\text{Ca})$ we find:
    \begin{align}
        a_2 &= \frac{3 \chi}{\chi^2 + \delta^2},
        &
        b_2 &= - \frac{1}{4} -\frac{1}{4} \frac{\chi}{\chi^2 + \delta^2},
        &
        c_2 &= -\frac{3 \delta}{\chi^2 + \delta^2},
        &
        d_2 &= \frac{1}{4} \frac{\delta}{\chi^2 + \delta^2},
        \\
        a_{-2} &= 2 - \frac{\chi}{\chi^2 + \delta^2},
        &
        b_{-2} &= \frac{1}{4} - \frac{1}{4} \frac{\chi}{\chi^2 + \delta^2},
        &
        c_{-2} &= -\frac{\delta}{\chi^2 + \delta^2},
        &
        d_{-2} &= -\frac{1}{4} \frac{\delta}{\chi^2 + \delta^2},
        \label{eqn:coeffs}
    \end{align}
    as well as $A_2 = 1$ and $B_2 = 0$,
    where we have defined 
    $\chi = (\mu^{E}_{D} + \mu^{E}_{M})/\mu^E_M = \lambda + 1$
    as the non-dimensionalised sum of the even viscosities 
    and
    $\delta = (\mu^{O}_{D} - \mu^{O}_{M})/\mu^E_M
    = \lambda \beta_D - \beta_M$
    as the non-dimensionalised difference of the odd viscosities.
    We immediately see that the velocity fields are determined by $\delta$, rather than by any of the odd-viscosity parameters individually.
    This feature is also observed in the case of flow past a translating droplet \cite{hosaka2021hydrodynamic}.
    In addition, we note that the droplet interface $\xi(\theta)$ is identical to the even case.
    Thus any dependence of the shape on odd viscosity must occur at order $\mathcal{O}(\text{Ca}^2)$ or higher.
    We also note that, in the rigid-particle limit $\chi \to \infty$, the coefficients become $b_2 \to -1/4$, $a_{-2} \to 2$, and $b_{-2} \to 1/4$, with all others vanishing.
    Therefore, the general solution may be decomposed as the sum of the rigid-particle solution and ``correction terms'', whose structure exhibits a high degree of mathematical symmetry.
    A gallery of analytical solutions is shown in Fig.~\ref{fig:pressure_Ca=0.01} for $\text{Ca} = 0.01$ and various values of the oddness parameters $\beta_D$ and $\beta_M$.
    The most striking feature is that, although the pressure fields have the same mathematical form in each case, their magnitudes are rescaled and their phases are shifted relative to the purely even case.
    The relative phases, $\phi_D$ and $\phi_M$, of the droplet and matrix pressure fields are plotted for $(\beta_M, \beta_D)$ in the region $[-1,1] \times [-1,1]$ of parameter space.
    The droplet phase shift $\phi_D$ decreases monotonically with both increasing $\beta_D$ and increasing $\beta_M$.
    Furthermore, $\phi_D = 0$ in the purely even case, as well as odd cases with $\delta = 0$. 
    The matrix phase shift $\phi_M$ increases monotonically with increasing $\beta_M$ but decreases monotonically with increasing $\beta_D$.
    Given the Lamb solution Eqs.~\eqref{eqn:lamb_u_r}-\eqref{eqn:lamb_p}, it should be possible to obtain analytical solutions for larger $\text{Ca}$ by including higher-order terms in the series expansion of $p$, $u_r$ and $u_\theta$ as well as the droplet shape $\xi$, in analogy with the three-dimensional treatments in \cite{BarthsBiesel1973, acrivos1983breakup, rallison1984deformation, Rallison_1980}.
    However, beyond leading order, such perturbative approaches are formidable opponent, so we turn to direct numerical simulations.

\clearpage
\subsection{Beyond Weak Shear Flows: Numerical Solutions}
    \label{sec:moderate_shear}
\begin{figure}[h]
        \centering
        \begin{subfigure}{\textwidth}
            \includegraphics[width=\linewidth]{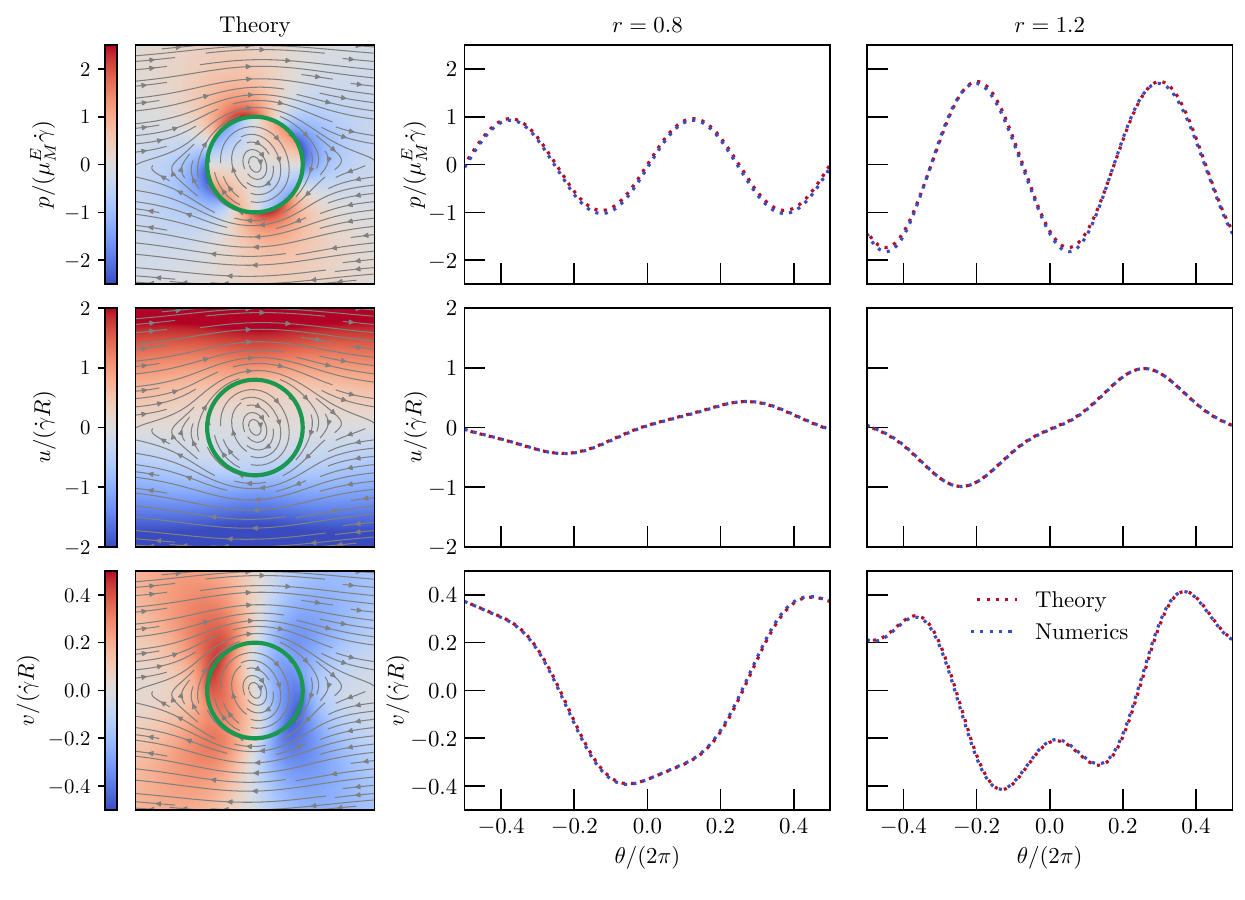}
        \end{subfigure}
        \caption{
            Comparison between numerical and analytical solutions for the flow around an odd-viscous droplet.
            Shown are the Cartesian velocity components $u$ and $v$, and the pressure $p$.
            The Laplace pressure, $\sigma/R$, has been subtracted from the droplet pressure field.
            The first column displays $u$, $v$, and $p$ with streamlines for the analytical solution Eqs.~\eqref{eqn:u_r_int}-\eqref{eqn:p_ext}.
            The second and third columns show $u$, $v$, and $p$ sampled along circles of radius $r = 0.8$ and $r = 1.2$, respectively, centred at the origin, for both numerical (Numerics) and analytical (Theory) solutions.
            For the numerical simulations, $\text{Re} = 0.01$, $\text{Ca} = 0.01$, and $L/R = 20$.
            For the analytical solutions, $\text{Re} = 0$, $\text{Ca} = 0$, and the domain is taken to be infinite.
            For both the analytical solution and numerical solutions, $\beta_D = 1.0$, $\beta_M = -1.0$.
        }
        \label{fig:DNScomparison}
    \end{figure}

    The results in this section are obtained from direct numerical simulations performed with Basilisk C~\cite{Popinet2009,Popinet2015}, using its Navier-Stokes and Volume-of-Fluid (VOF) solvers together with the odd-viscous stress module developed in \cite{francca2025odd}.
    All simulations are carried out at low Reynolds number, $\text{Re} = 0.01$, so that the steady-state numerical solutions provide a good approximation to creeping flow.
    Appendix~\ref{appendix:numerics} presents a more detailed account of the numerical method, including the full equations of motion, the simulation initialisation, transient dynamics, and a comprehensive comparison with the analytical solution in Section~\ref{sec:weak_shear}.      
    In Fig.~\ref{fig:DNScomparison}, we present one such example comparison of the steady-state numerical solution and the analytical solution for a weak shear case ($\text{Ca} = 0.01$) with dimensionless odd viscosities $\beta_D = 1.0$ and $\beta_M = -1.0$.

    \begin{figure}[h]
        \centering
        \includegraphics[width=\linewidth]{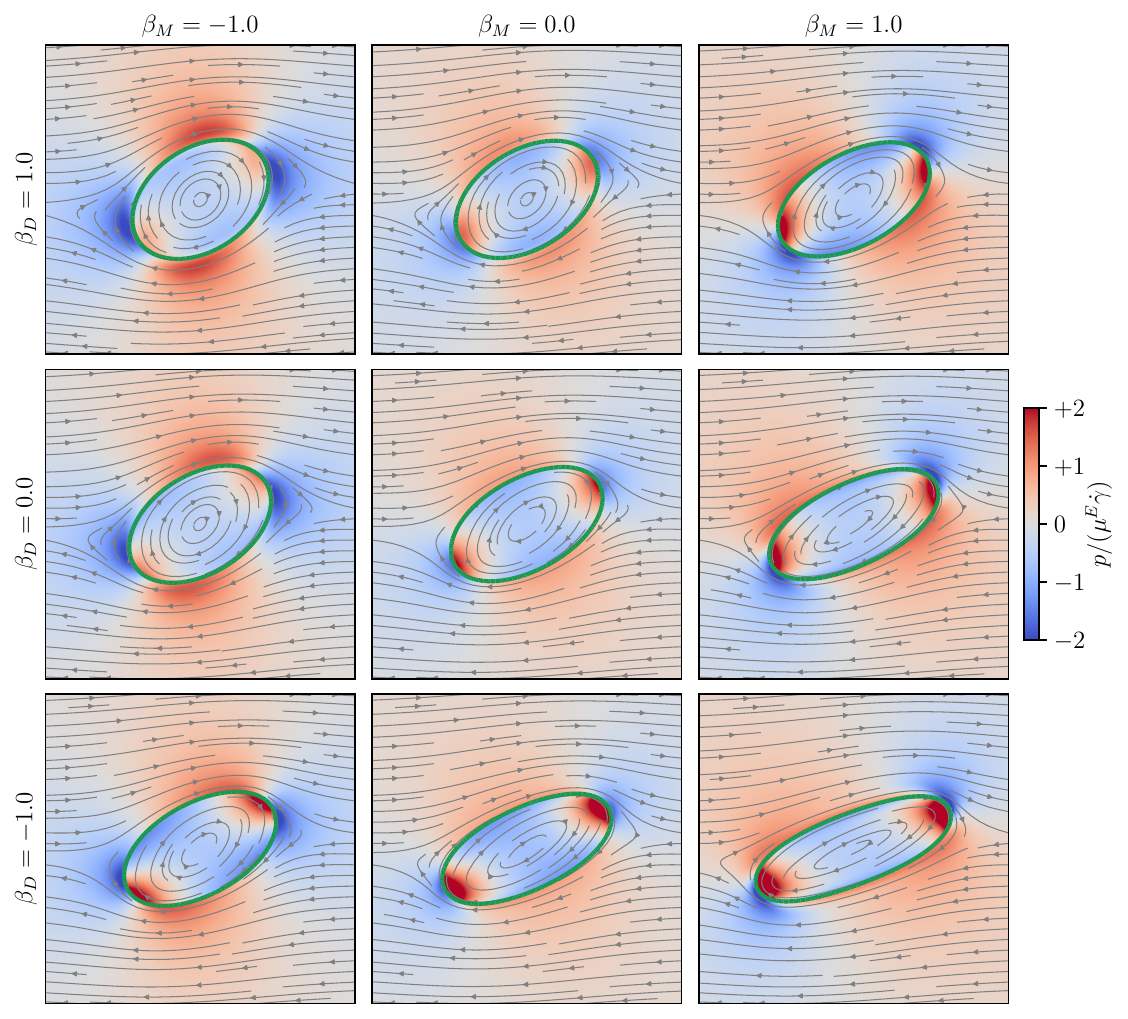}
        \caption{
            The pressure field $p$ for an odd-viscous droplet in an odd-viscous simple shear flow according to direct numerical simulations.
            The Laplace pressure, $\sigma/R$, has been subtracted from the droplet pressure field.
            The parameters are $\text{Re} = 0.01$, $\text{Ca} = 0.3$, $\lambda = 1$ and $L/R = 20$.
            Panels show different combinations of the oddness parameters $(\beta_D, \beta_M) \in [-1, 1] \times [-1, 1]$.
            The plotted region is $[-2.5R, 2.5R] \times [-2.5R, 2.5R]$.
            The grey curves denote streamlines of the velocity field $\bm{u}$.
        }
        \label{fig:deform_pressure_Ca=0.3}
    \end{figure}
    
    \begin{figure}[h]
        \centering
        \begin{subfigure}{0.32\textwidth}
            \includegraphics[width=\linewidth]{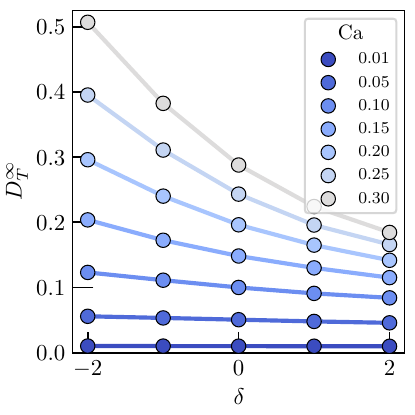}
            \subcaption{}
        \end{subfigure}
        \begin{subfigure}{0.32\textwidth}
            \includegraphics[width=\linewidth]{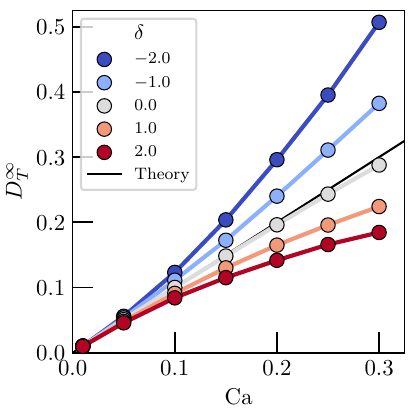}
            \subcaption{}
        \end{subfigure}
        \begin{subfigure}{0.32\textwidth}
            \includegraphics[width=\linewidth]{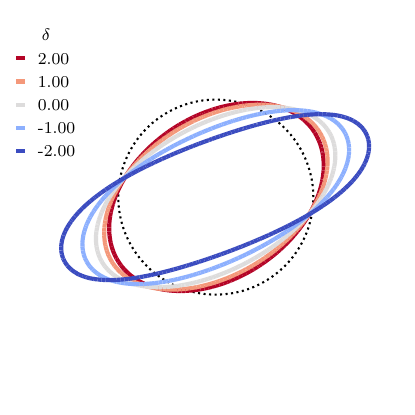}
            \subcaption{}
        \end{subfigure}
        \caption{
            Effect of odd viscosity on steady-state droplet deformation.
            (a) $D_T^\infty$ as a function of $\delta$ for different values of $\text{Ca}$.
            (b) $D_T^\infty$ as a function of $\text{Ca}$ for different values of $\delta$.
            The solid black line indicates the leading-order prediction $D_T^\infty = \text{Ca} + \mathcal{O}(\text{Ca}^2)$.
            (c) Steady-state droplet interface shapes for $\text{Ca} = 0.3$ and different values of $\delta$.
        }
        \label{fig:deformation_theory}
    \end{figure}

    Fig.~\ref{fig:deform_pressure_Ca=0.3} shows numerical solutions for the flow around droplets in moderate shear ($\text{Ca} = 0.3$) in steady state, for a fixed even viscosity ratio $\lambda = 1$ and a range of dimensionless odd viscosities $-1.0 < \beta_D < 1.0$ and $-1.0 < \beta_M < 1.0$.
    For all $\beta_D$ and $\beta_M$ the flow remains qualitatively similar, but the angle subtended by the local extrema of the pressure depends on $\beta_D$ and $\beta_M$.
    Simulations with the same value of $\delta$ exhibit identical droplet shapes and velocity fields.
    Moreover, the degree of droplet deformation decreases monotonically with $\delta$: simulations with negative $\delta$ produce more strongly deformed droplets whose long axis lies closer to the horizontal, whereas those with positive $\delta$ yield more weakly deformed droplets whose long axis is more diagonal.
    
    Fig.~\ref{fig:deformation_theory}(a) shows $D_T^\infty$ as function of $\delta$.
    Fig.~\ref{fig:deformation_theory}(b) shows $D_T^\infty$ as a function of $\text{Ca}$.
    Shown in Fig.~\ref{fig:deformation_theory}(c) is a comparison of the droplet interfaces in steady state for different $\delta$.
    It follows from the solution in Section~\ref{sec:weak_shear} that $D_T^\infty = \text{Ca} + \mathcal{O}({\text{Ca}}^2)$.
    This result is shown with a solid black line in Fig.~\ref{fig:deformation_theory}(b).
    It turns out, (see Appendix~\ref{appendix:steady_state}), that $D_T^\infty$ and the steady-state orientation angle $\theta^\infty$ are correlated in a way that is independent of oddness, which strongly suggests that the interfaces in Fig.~\ref{fig:deform_pressure_Ca=0.3} belong to the same family of droplet shapes. 
    This indicates that, for weak and moderate shear flows at least, the addition of odd viscosity does not lead to a different set of droplet shapes compared to the purely even case.
    For a more detailed discussion of the evidence that the flow and shape only depend on $\delta$, readers are advised to read Appendix~\ref{appendix:numerics}.
    Appendix~\ref{appendix:numerics} also briefly discusses larger deformations at larger values of $\text{Ca}$.

\clearpage
\section{The Droplet in Shear as a Dilute Odd Emulsion}
    \label{sec:apparent_viscosity}
    \begin{figure}[h]
        \centering
        \begin{subfigure}{0.49\linewidth}
            \includegraphics[width=\linewidth]{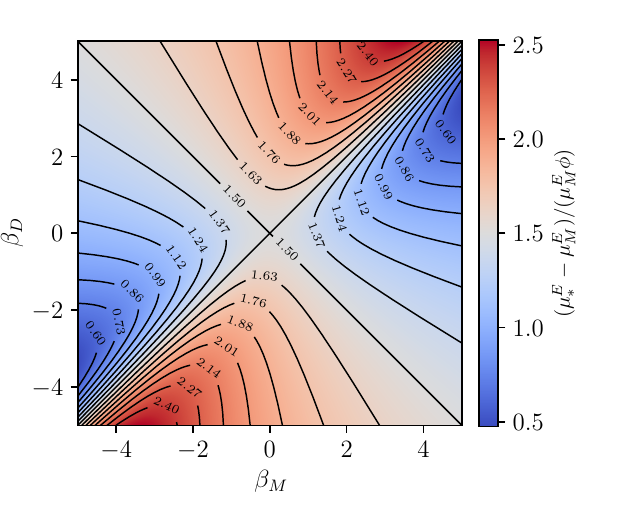}
            \caption{}
            \label{fig:mu_app_even_map}
        \end{subfigure}
        \begin{subfigure}{0.49\linewidth}
            \includegraphics[width=\linewidth]{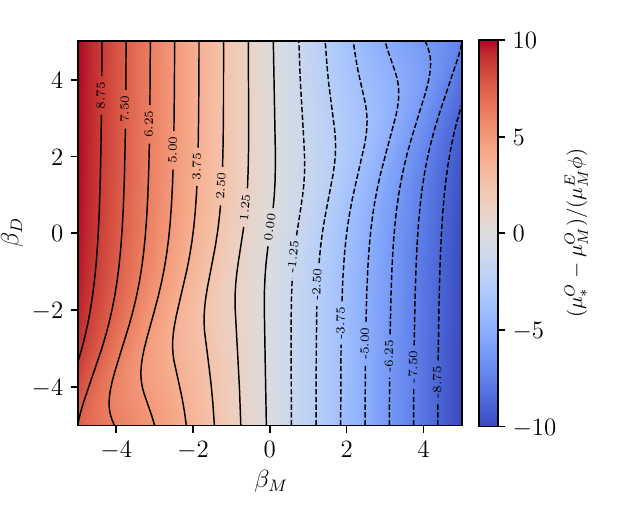}
            \caption{}
            \label{fig:mu_app_odd_map}
        \end{subfigure}
        \begin{subfigure}{0.49\linewidth}
            \includegraphics[width=\linewidth]{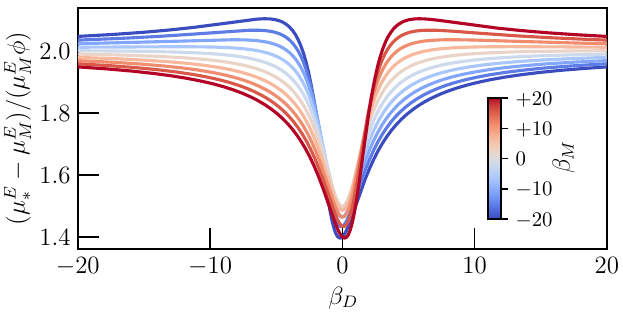}
            \caption{}
            \label{fig:mu_app_even_graph}
        \end{subfigure}
        \begin{subfigure}{0.49\linewidth}
            \includegraphics[width=\linewidth]{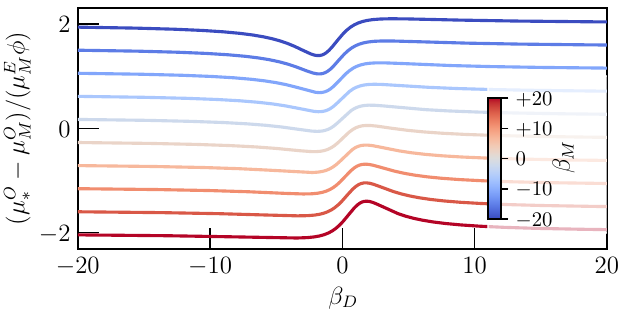}
            \caption{}
            \label{fig:mu_app_odd_graph}
        \end{subfigure}
        
        \caption{
            Apparent even and odd viscosities of a dilute droplet suspension from the analytical expression 
            Eqs.~\eqref{eqn:mu_app_even}–\eqref{eqn:mu_app_odd}.
            Shown in (a) and (b) are the relative changes $(\mu_*^E - \mu_M^E)/(\mu_M^E \phi)$ and
            $(\mu_*^O - \mu_M^O)/(\mu_M^E \phi)$ respectively, as functions of the oddness parameters
            $(\beta_M, \beta_D)$ in the region $[-5,5] \times [-5,5]$, for $\lambda = 1$.
            These quantities represent the dimensionless change in even and odd viscosity of the system respectively due to the droplet, measured in units of the area fraction $\phi$ covered by the droplet.
            Contour lines are superimposed.
            Colour maps are centred on the purely even case $\beta_D = \beta_M = 0$.
            Shown in (c) and (d) are those same quantities for fixed values of $\beta_M \in [-20,20]$
        }
        \label{fig:apparent_viscosity}
    \end{figure}
    Given the analytical solution in Section III.\ref{sec:weak_shear} for the flow around a single droplet under shear, we can derive an expression for the effective viscosity of a dilute emulsion, in the limit of weak shear.
    In the dilute limit, the bulk stress 
    $\bm{\tau}_*$
    of the emulsion is
    $\bm{\tau}_* = 2\mu_M^E \bm{E}^\infty + \mu_M^O(\bm{\epsilon}\cdot\bm{E}^\infty - \bm{E}^\infty \cdot\bm{\epsilon}) + n \bm{S}$
    where 
    $\bm{E}^\infty$
    is the undisturbed strain-rate tensor, $n$ is the number of droplets per unit area, and 
    $\bm{S}$
    is the symmetric part of the first moment of the stress distribution 
    $\bm{\sigma}_M$
    around a droplet, also known as the ``stresslet''.
    The stresslet can be computed via the following line integral over the droplet interface:
    \begin{align}
        \bm{S}
        &=
        \frac{1}{2}
        \oint_{\xi} 
        \dd{l}
        \left[
            (\bm{\sigma}_M\cdot \bm{\hat{n}}) \otimes \bm{x}
            +
            \bm{x} \otimes (\bm{\sigma}_M\cdot \bm{\hat{n}}) 
            -
            (\bm{x} \cdot \bm{\sigma}_M \cdot \bm{\hat{n}})
            \,\bm{\delta}
        \right].
    \end{align} 
    where $\bm{\delta}$ is the identity tensor.
    This may be computed directly for the analytical solution in Section III.\ref{sec:weak_shear} to find:
    \begin{align}
        \bm{S}
        &=
        \pi R^2
        \left[
            2(a_{-2} \mu_M^E - c_{-2}\mu_M^O) \bm{E}^\infty
            -
            (a_{-2} \mu_M^O + c_{-2}\mu_M^E) (\bm{\epsilon} \cdot \bm{E}^\infty - \bm{E}^\infty \cdot \bm{\epsilon})
        \right].
    \end{align} 
    We define the apparent even and odd viscosities, $\mu_*^E$ and $\mu_*^O$, as the coefficients in the constitutive relation
    $\bm{\tau}_* = 2\mu_*^E \bm{E}^\infty + \mu_*^O(\bm{\epsilon} \cdot \bm{E}^\infty - \bm{E}^\infty \cdot \bm{\epsilon})$
    so that $\mu_*^E$ and $\mu_*^O$ may be interpreted as the even and odd viscosities of an effective homogeneous fluid with the same macroscopic shear response as the droplet suspension.
    It follows immediately that $\mu_*^E$ and $\mu_*^O$ are given by:
    \begin{align}
        \mu_*^E
        &=
        \mu_M^E 
        +
        \left(
            a_{-2} \mu_M^E - c_{-2}\mu_M^O
        \right) \phi,
        \label{eqn:mu_app_even}
        \\
        \mu_*^O
        &=
        \mu_M^O
        -
        \left(
            a_{-2} \mu_M^O + c_{-2}\mu_M^E
        \right) \phi,
        \label{eqn:mu_app_odd}
    \end{align}
    where $\phi$ is the area fraction covered by the droplet phase, and $a_{-2}$ and $c_{-2}$ are the constants stated in Eq.~\eqref{eqn:coeffs}.
    Eqs.~\eqref{eqn:mu_app_even} and~\eqref{eqn:mu_app_odd} are plotted in Fig.~\ref{fig:apparent_viscosity}, for even viscosity ratio $\lambda = 1$.
    We choose to plot the dimensionless differences, $(\mu_*^E - \mu_M^E)/\mu_M^E$ and $(\mu_*^O - \mu_M^O)/\mu_M^E$,
    between the bulk viscosities and matrix viscosities, measured in units of the area fraction $\phi$ of the droplet.
    In this way, Eqs.~\eqref{eqn:mu_app_even} and~\eqref{eqn:mu_app_odd} may be plotted without choosing an arbitrary value of $\phi$.
    Since the system is assumed dilute, however, $\phi \ll 1$ and the relative changes in viscosity are always small compared to the viscosities of the bulk.
    When $\lambda = 1$ and $\delta = 0$, $\mu_*^E = \mu_M^E(1 + 3\phi/2)$ and $\mu_*^O = 0$.
    These value are used to centre the colour maps in Figs.~\ref{fig:mu_app_even_map} and~\ref{fig:mu_app_odd_map}.
    From Fig.~\ref{fig:mu_app_even_map}, it is apparent that whenever $\beta_D = \pm \beta_M$, the value of $\mu_*^E$ is identical to the purely even case.
    This contour divides the plot into four quadrants of increased and decreased $\mu_*^E$ relative to the purely even case.
    Fig.~\ref{fig:mu_app_even_graph} shows a series of vertical slices through Fig.~\ref{fig:mu_app_even_map}.
    It is apparent that for $\abs{\beta_D}\to\infty$, $\mu_*^E \to \mu_M^E(1 + 2\phi)$, which is precisely the rigid particle limit.
    It is also clear that the minimum value of $\mu_*^E$ occurs for small values of $\beta_D$.
    This indicates that a dilute suspension of highly odd droplets, be they positive or negative in odd viscosity, leads to an increase in bulk energy dissipation, as though the droplets were instead rigid particles.
    Odd viscosity, though non-dissipative at the micro scale, leads to a net increase in dissipation in the bulk relative to the equivalent system of droplets with only even viscosity.
    Fig.~\ref{fig:mu_app_odd_map} shows that $\mu_*^O$ does not depend strongly on $\beta_D$.
    In addition, it shows that, except when $\beta_M$ is small, the effect of adding droplets to the matrix, regardless the value of $\beta_D$, is to bring $\mu_*^O$ closer to zero.
    That is to say, when $\beta_M$ is positive, the contribution of the droplets is negative and vice versa.
    Fig.~\ref{fig:mu_app_even_graph} shows a series of vertical slices through Fig.~\ref{fig:mu_app_even_map}.
    It is apparent that the limits $\beta_D\to\pm\infty$ are identical, with non-monotonicities near $\beta_D = 0$.
    This behaviour largely rules out the possibility of increasing the relative oddness of the matrix by the addition of droplets, except in cases where $\beta_M$ is sufficiently small.
        
\section{Conclusion}
    We have derived an analytical solution to the droplet in shear problem in the setting of two-dimensional odd Stokes flow in the limit $\text{Ca} \ll 1$.
    From this, we have obtained expressions for the apparent even and odd viscosities of a dilute emulsion,
    $
    \mu_*^E
    =
    \mu_M^E 
    +
    (a_{-2} \mu_M^E - c_{-2}\mu_M^O) \phi
    $
    and
    $
    \mu_*^O
    =
    \mu_M^O
    -
    (a_{-2} \mu_M^O + c_{-2}\mu_M^E) \phi
    $
    where
    $a_{-2} = 2 - \chi/(\chi^2 + \delta^2)$
    ,
    $c_{-2} = - \delta/(\chi^2 + \delta^2)$
    ,
    $\chi = (\mu^{E}_{D} + \mu^{E}_{M})/\mu^E_M$
    ,
    $\delta = (\mu^{O}_{D} - \mu^{O}_{M})/\mu^E_M$
    and $\phi$ is the area fraction of the dispersed phase.
    We have shown that to leading order, the steady-state Taylor deformation parameter is 
    $D_T^\infty = \text{Ca} + \mathcal{O}({\text{Ca}}^2)$
    independent of the even viscosity ratio and odd viscosities, provided none of these parameters is too large.
    Furthermore, the influence of odd viscosity on the velocity field enters only through the difference
    $\delta = (\mu^{O}_{D} - \mu^{O}_{M})/\mu^E_M$
    rather than on $\mu_M^O$ and $\mu_D^O$ separately. 
    For larger values of $\text{Ca}$, we have performed numerical simulations which indicate that a positive odd viscosity difference $\delta > 0$ leads to a decrease in steady state droplet deformation $D_T^\infty$ and vice versa.
    However, the simulations also suggest that regardless the values of $\delta$ and $\text{Ca}$, $D_T^\infty$ and the steady-state orientation angle $\theta^\infty$ are correlated in a way that is independent of $\delta$, suggesting that in spite of odd viscosity, the steady-state interfaces belong to the same family of curves.
    To confirm definitively that this is in indeed the case, a more careful analysis is needed, perhaps involving a higher-order perturbation theory.
    For much larger $\text{Ca}$ ($\text{Ca} = 1.0$), the numerical simulations suggest tentative evidence of a breakdown in $\delta$-universality, with the shape of the droplet and thus also the velocity fields showing some dependence on the individual odd viscosities.
    However, definitive proof of the existence of the breakdown would ultimately rest on a formal mathematical argument, which we have not been able to find.

    Our results may have implications for the measurement of odd viscosity in experiments.
    For instance, regarding the quasi-two-dimensional chiral fluid composed of spinning colloidal particles (\emph{e.g.,} those in Soni \emph{et al}.~\cite{Soni2019}), one could envision a system in which both magnetic and non-magnetic particles were introduced (or particles with different magnetic moments) to form phases with different chiralities.
    The resulting system, under certain assumptions, might then be approximated by the present theoretical model and its bulk apparent viscosity given by the present formulas.
    In principle, one could then extract information each phase from the slope of the measured apparent even and odd viscosity as a function of the area fraction $\phi$.
    This is the odd-viscosity analogue of the classical Einstein/Taylor viscometry relations, and represents an indirect method for characterising chiral fluids that does not require resolving the microscopic flow field.

    More broadly, our results contribute to a growing understanding of how chirality arises across scales.
    It is already understood how parity-broken interactions between the microscopic constituents lead to a parity-broken continuum description of a chiral fluid.
    In our work, we take steps towards showing how mesoscopic chiral droplets contribute to a course-grained description of the bulk.
    By considering a simple geometry and an unconfined domain, we obtain exact results for the even and odd viscosity of a dilute emulsion.
    The framework developed here, is transferable to other mesoscale geometries including vesicles, active droplets, and chiral inclusions in elastic media, and we hope it will serve as a building block for the rheology of more complex heterogeneous chiral soft matter systems.
    Natural immediate extensions of our work include the effects of confinement, droplet-droplet interactions in non-dilute emulsions, and also exploring the role of the even viscosity ratio $\lambda$.

\clearpage
\appendix
\textbf{\Large Appendices}

\section{Numerical Simulations}
    \label{appendix:numerics}
    In this appendix, we describe the implementation of the direct numerical simulations (DNS) used to generate the solutions presented in Sections II.~\ref{sec:moderate_shear} and III.~\ref{sec:strong_shear}.
    The DNS are performed with Basilisk C, an open-source language developed by S.~Popinet and collaborators~\cite{Popinet2003, Popinet2009, Popinet2015} for solving differential equations on adaptive Cartesian meshes. 
    In particular, we use the built-in Navier-Stokes and Volume-of-Fluid (VOF) solvers, which have been extensively validated in previous studies \cite[see][for detailed comparisons]{Popinet2009}, together with the implementation of odd-viscous stresses introduced in~\cite{francca2025odd} (also see~\cite{duPlessis2025}).
    
    The maximum level of mesh refinement, \texttt{LEVEL}, is defined such that the domain size $L$ is $2^{\texttt{LEVEL}}$ times the size of the smallest cell.
    Consequently, the maximum number of cells across the initial droplet diameter is
    $2^{\texttt{LEVEL}} \cdot 2R/L$,
    where $R$ is the initial droplet radius.
    Throughout this work, we choose $L/R = 20$ so that interactions between the droplet and the domain boundaries can be neglected.
    Unless otherwise stated, we use \texttt{LEVEL} $= 9$, corresponding to a maximum of roughly 50 cells per droplet diameter.
    For the weak-shear case $\text{Ca} = 0.01$, a resolution of \texttt{LEVEL} $= 11$ is necessary to accurately resolve the deformation parameter $D_T$.

\subsection*{Initialisation}
    \label{sec:initialisation}
    \begin{figure}[h]
        \centering
        \includegraphics[width=0.45\linewidth]{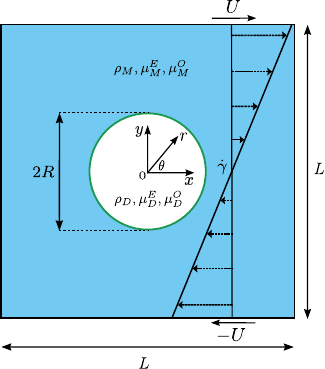}
        \caption{
            Initial configuration of the droplet under shear DNS.
            A single droplet of radius $R$ is located at the centre $O$ of a square domain of side length $L$. 
            The droplet has density $\rho_D$, even viscosity $\mu_D^E$ and odd viscosity $\mu_D^O$.
            The matrix has density $\rho_M$, even viscosity $\mu_M^E$ and odd viscosity $\mu_M^O$.
            The upper and lower boundaries move with speed $U = \dot{\gamma} L/2$ in opposite directions, yielding an average shear rate $\dot{\gamma}$.
            Periodic boundary conditions are imposed on the left and right edges.
            The surface tension at the interface is $\sigma$.
        }
        \label{fig:configuration}
    \end{figure}
    The initial configuration is depicted in Fig.~\ref{fig:configuration}. We consider a single droplet of radius $R$ located at the centre $O$ of a square domain $\Omega$ of side length $L$. We adopt a Cartesian coordinate system $\bm{x} = (x,y)$ centred on $O$, with axes parallel to the boundaries of $\Omega$. With this choice, the domain is $\Omega = [-L/2, +L/2] \times [-L/2, +L/2]$, and the droplet centre is at $\bm{0} = (0,0)$.
    We use subscripts $_D$ and $_M$ to refer to the droplet and matrix, respectively. The fluid in the droplet interior $\Omega_D$ has density $\rho_D$, even viscosity $\mu_D^E$, and odd viscosity $\mu_D^O$, while the exterior region $\Omega_M$ has density $\rho_M$, even viscosity $\mu_M^E$, and odd viscosity $\mu_M^O$. 
    At the droplet interface $S = \partial \Omega_D$ we assume a constant surface tension coefficient $\sigma$.
    
    At time $t = 0$, the velocity field is initialised everywhere to be the undisturbed simple shear flow $\bm{u}^\infty = (\dot{\gamma} y, 0)$, where $\dot{\gamma}$ is the imposed shear rate, so that the entire domain is immediately in motion. 
    The upper and lower boundaries of the domain are interpreted as parallel plates, each moving with speed $U = \dot{\gamma} L/2$ in opposite directions, yielding an average shear rate $\dot{\gamma}$. 
    We therefore impose no-slip boundary conditions $\bm{u}(x, \pm L/2) = (\pm U, 0)$.
    In addition, we impose periodic boundary conditions on the left and right edges such that the configuration is equivalent to an infinite one-dimensional array of droplets with uniform spacing $L$.

\subsection*{Equations of Motion}
    The governing equations are the incompressible odd Navier-Stokes equations,
    \begin{align}
        \bm{\nabla} \cdot \bm{u} &= 0
        \label{eqn:continuity_basilisk}
        \\
        \rho
        \left(\pdv{\bm{u}}{t} + \bm{u}\cdot\bm{\nabla u}\right)
        &=
        -\bm{\nabla} p
        +
        \bm{\nabla} \cdot 
        \left(
            2\mu^E \bm{E} + \mu^O[\bm{\epsilon}\cdot\bm{E} - \bm{E}\cdot\bm{\epsilon}]
        \right)
        +
        \bm{f}_\sigma.
        \label{eqn:momentum_basilisk}
    \end{align}
    where $\rho$ is the mass density, $\mu^E$ and $\mu^O$ are the even and odd viscosities, $\bm{E}$ is the rate-of-strain tensor, and $\omega$ is the vorticity. 
    The disparate material properties of the droplet and matrix are encoded using a scalar colour function $c(\bm{x},t)$, which takes the value 0 inside $\Omega_D$ and 1 inside $\Omega_M$. The density field is then
    $\rho = \rho_M c(\bm{x},t) + \rho_D[1 - c(\bm{x},t)]$, and the even and odd viscosities are
    $\mu^{E,O} = \mu_M^{E,O} c(\bm{x},t) + \mu_D^{E,O}[1 - c(\bm{x},t)]$. 
    The time evolution of $\Omega_D$ and $\Omega_M$ is obtained by advecting $c(\bm{x},t)$ with the velocity field $\bm{u}$ using the VOF method. 
    The surface-tension force $\bm{f}_\sigma$ in the momentum equation is written as
    $\bm{f}_\sigma = \sigma \kappa \bm{\hat{n}} \delta_S$, where $\kappa$ is the mean curvature of the interface, $\bm{\hat{n}}$ its unit normal, and $\delta_S$ a Dirac delta distribution restricting the force to the interface $S$.
    Equations~\eqref{eqn:continuity_basilisk} and~\eqref{eqn:momentum_basilisk} are nondimensionalised using the initial droplet radius $R$, the imposed average strain rate $\dot{\gamma}$, and the matrix even shear-stress scale $\mu_M^E \dot{\gamma}$, yielding
    \begin{align}
        \bar{\bm{\nabla}} \cdot \bar{\bm{u}} &= 0
        \label{eqn:continuity_basilisk_nondim}
        \\
        \bar{\rho}
        \text{Re}
        \left(\pdv{\bar{\bm{u}}}{\bar{t}} + \bar{\bm{u}} \cdot \bar{\bm{\nabla}} \bar{\bm{u}}\right)
        &=
        -\bar{\bm{\nabla}} \bar{p}
        +
        \bar{\bm{\nabla}} 
        \cdot 
        \left(
            2\bar{\mu}^E\bar{\bm{E}}
            + 
            \bar{\mu}^O[\bm{\epsilon}\cdot\bar{\bm{E}} - \bar{\bm{E}}\cdot\bm{\epsilon}]
        \right)
        +
        \frac{1}{\text{Ca}}
        \bar{\kappa} \bm{\hat{n}} \bar{\delta}_S
        \label{eqn:momentum_basilisk_nondim}
    \end{align}
    where the $\bar{}$ symbol indicates a quantity is dimensionless.
    Here
    \begin{align}
        \bar{\rho} &= c(\bar{\bm{x}},\bar{t}) + \alpha[1 - c(\bar{\bm{x}},\bar{t})] \\
        \bar{\mu}^{E} &= c(\bar{\bm{x}},\bar{t}) + \lambda [1 - c(\bar{\bm{x}},\bar{t})] \\
        \bar{\mu}^{O} &= \beta_M c(\bar{\bm{x}},\bar{t}) + \lambda\beta_D [1 - c(\bar{\bm{x}},\bar{t})] 
    \end{align}
    where 
    $\alpha = \rho_D/\rho_M$ is the density ratio,
    $\lambda = \mu_D^E/\mu_M^E$ is the even viscosity ratio 
    and
    $\beta_D = \mu_D^O/\mu_D^E$ and $\beta_M = \mu_M^O/\mu_M^E$ are the ``oddness" parameters of the droplet and matrix respectively.
    The other dimensionless groups that appear are the even Reynolds number
    $\text{Re} = \rho_M \dot{\gamma} R^2/\mu_M^E$ 
    and the even capillary number
    $\text{Ca} = \mu_M^E \dot{\gamma} R/\sigma$.
    It is also convenient to introduce the additional dimensionless parameter $\phi = \pi R^2 / L^2$, which measures the area fraction of the domain $\Omega$ occupied by the droplet region $\Omega_D$.
    Throughout this work, we fix $\alpha = 1$ and $\lambda = 1$ such that each phase has equal density and equal even viscosity.
    We also fix $L/R = 20$ such that $\phi \approx 0.008$, which makes the system dilute.
    The even Reynolds number is set to $\text{Re} = 0.01$ to approximate the creeping-flow regime. 
    The remaining control parameters are the even capillary number $\text{Ca}$ and the oddness parameters $\beta_D$ and $\beta_M$.

\clearpage
\section{Transient Dynamics}
    \label{appendix:time-dependence}
\subsection*{Weak to Moderate Shear Flows}
    \begin{figure}[h]
        \centering
        \includegraphics[width=\textwidth]{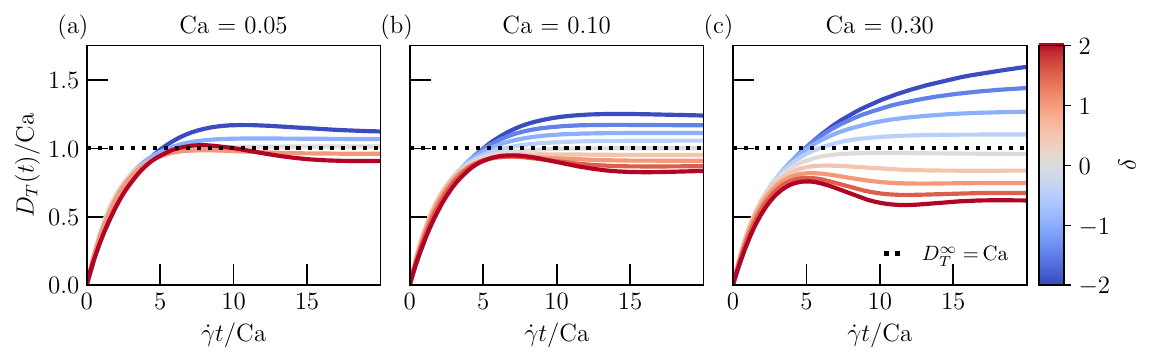}
        \caption{
            The effect of odd viscosity on time-dependent droplet deformation.
            All simulations use $\text{Re} = 0.01$, $\lambda = 1$ and $L/R = 20$.
            Shown are the evolutions of the Taylor deformation parameter $D_T(t)$ for
            (a) $\text{Ca} = 0.05$,
            (b) $\text{Ca} = 0.1$, and
            (c) $\text{Ca} = 0.3$,
            each for several values of the odd-viscosity difference $\delta$.
            The black dotted line shows the leading order theory
            $D_T^\infty = \text{Ca} + \mathcal{O}({\text{Ca}}^2)$.
        }
        \label{fig:time_dep_deform}
    \end{figure}

\begin{figure}[h]
        \centering
        \begin{subfigure}{0.3\textwidth}
            \includegraphics[width=\linewidth]{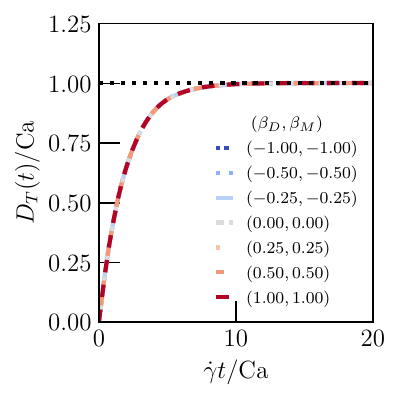}
            \subcaption{}
        \end{subfigure}
        \begin{subfigure}{0.3\textwidth}
            \includegraphics[width=\linewidth]{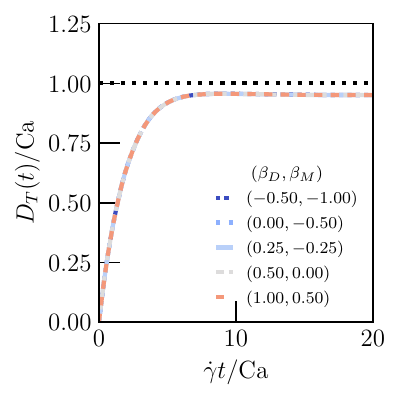}
            \subcaption{}
        \end{subfigure}
        \begin{subfigure}{0.3\textwidth}
            \includegraphics[width=\linewidth]{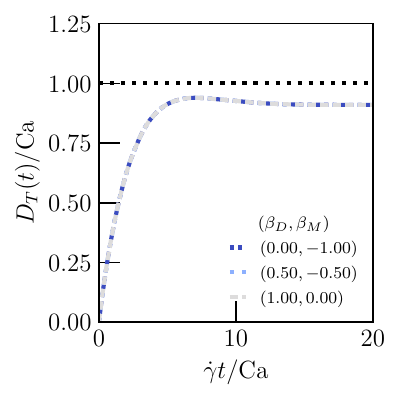}
            \subcaption{}
        \end{subfigure}
        \begin{subfigure}{0.3\textwidth}
            \includegraphics[width=\linewidth]{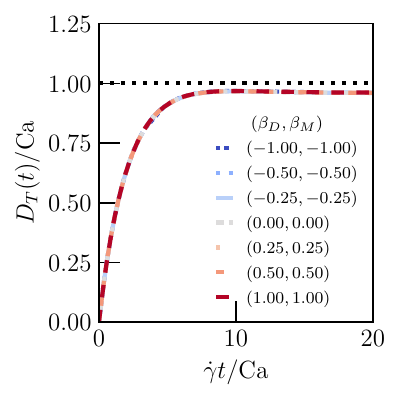}
            \subcaption{}
        \end{subfigure}
        \begin{subfigure}{0.3\textwidth}
            \includegraphics[width=\linewidth]{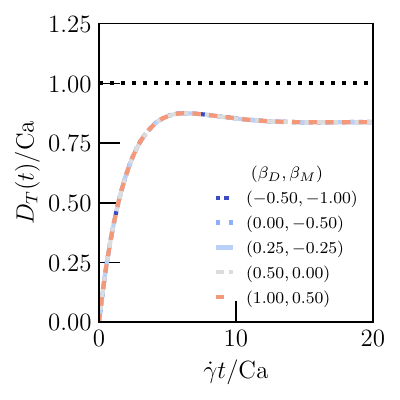}
            \subcaption{}
        \end{subfigure}
        \begin{subfigure}{0.3\textwidth}
            \includegraphics[width=\linewidth]{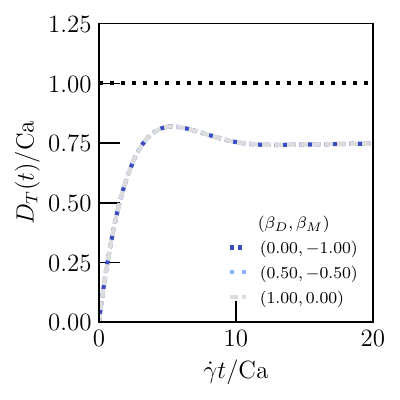}
            \subcaption{}
        \end{subfigure}
        \caption{
            The effect of the oddness parameters $\beta_D$ and $\beta_M$ on the time-dependent droplet deformation.
            All simulations use $\text{Re} = 0.01$ $\lambda = 1$ and $L/R = 20$.
            Shown are the evolutions of the Taylor deformation parameter $D_T(t)$ for
            (a) $\delta = 0.0, \text{Ca} = 0.1$,
            (b) $\delta = 0.5, \text{Ca} = 0.1$, 
            (c) $\delta = 1.0, \text{Ca} = 0.1$,
            (d) $\delta = 0.0, \text{Ca} = 0.3$,
            (e) $\delta = 0.5, \text{Ca} = 0.3$
            and 
            (f) $\delta = 1.0, \text{Ca} = 0.3$,
            each for several combinations $\beta_D$ and $\beta_M$ consistent with the chosen value of $\delta$.
            The black dotted line shows the leading order theory
            $D_T^\infty = \text{Ca} + \mathcal{O}({\text{Ca}}^2)$.
        }
        \label{fig:time_dep_deform_delta_dependence}
    \end{figure}

    Fig.~\ref{fig:time_dep_deform} shows the time dependence of the deformation parameter $D_T(t)$ for fixed values of $\text{Ca}$, namely $\text{Ca} = 0.05$, $0.1$, and $0.3$, each with various values of the odd viscosity difference $\delta = (\mu_D^O - \mu_M^O)/\mu_M^E$ in the range $-2.0 \leq \delta \leq 2.0$. 
    Different pairs of $\mu_D^O$ and $\mu_M^O$ values with equal $\delta$ give identical dynamics.
    This is demonstrated in Fig.~\ref{fig:time_dep_deform_delta_dependence} for $\text{Ca} = 0.1$ and $0.3$ for $\delta = 0, 0.5$ and $1.0$.
    In each simulation, the droplet begins to deform immediately at $t = 0$, and $D_T$ increases from its initial value. 
    In each case, the droplet eventually reaches a steady state in which $D_T$ approaches an equilibrium value $D_T^\infty$ on a timescale that grows with $\text{Ca}$. 
    The steady value is known to satisfy $D_T^\infty = \text{Ca} + \mathcal{O}({\text{Ca}}^2)$ when none of the viscosity ratios is too large.
    For large positive values of $\delta$, $D_T(t)$ overshoots its equilibrium value, indicating underdamped dynamics. 
    The degree of damping decreases monotonically as $\delta$ increases. 
    In the DNS, we regard the system as having reached steady state when both $D_T$ and the total kinetic energy in each phase have saturated to time-independent values.

\clearpage
\subsection*{Strong Shear Flows}
    \label{sec:strong_shear}
        \begin{figure}[h]
        \centering
        \begin{subfigure}{0.3\textwidth}
            \includegraphics[width=\linewidth]{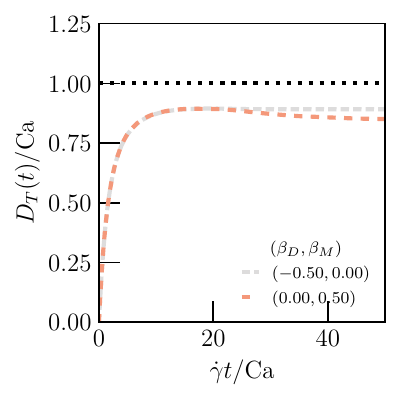}
            \subcaption{}
        \end{subfigure}
        \begin{subfigure}{0.3\textwidth}
            \includegraphics[width=\linewidth]{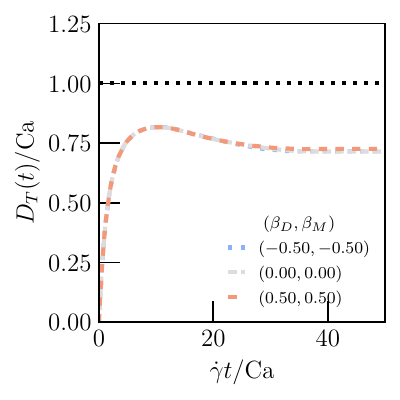}
            \subcaption{}
        \end{subfigure}
        \begin{subfigure}{0.3\textwidth}
            \includegraphics[width=\linewidth]{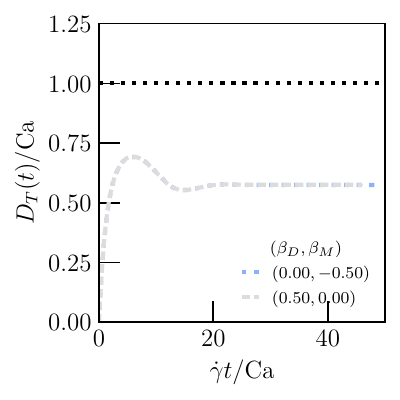}
            \subcaption{}
        \end{subfigure}
        \caption{
            The effect of the oddness parameters $\beta_D$ and $\beta_M$ on the time-dependent droplet deformation.
            All simulations use $\text{Re} = 0.01$, $\lambda = 1$, $\text{Ca} = 1.0$ and $L/R = 20$.
            Shown are the evolutions of the Taylor deformation parameter $D_T(t)$ for
            (a) $\delta = -0.5$,
            (b) $\delta = 0.0$
            and
            (c) $\delta = 0.5$,
            each for at least two combinations $\beta_D$ and $\beta_M$ consistent with the chosen value of $\delta$.
            The black dotted line shows the leading order theory
            $D_T^\infty = \text{Ca} + \mathcal{O}({\text{Ca}}^2)$.
        }
        \label{fig:time_dep_deform_delta_dependence_strong_shear}
    \end{figure}
    Fig.~\ref{fig:time_dep_deform_delta_dependence_strong_shear} shows $D_T(t)$ at $\text{Ca} =1.0$ and $\delta = -0.5, 0$ and $0.5$.
    For each value of $\delta$ there are at least two combinations $\beta_D$ and $\beta_M$ consistent with the chosen value of $\delta$.
    For $\delta = 0$ and $0.5$, the dynamics are independent of $\beta_D$ and $\beta_M$, just as for the weak shear cases.
    However, for $\delta = -0.5$ the combination $(\beta_D, \beta_M) = (0.0, 0.5)$ leads to a smaller value of $D_T^\infty$ than $(\beta_D, \beta_M) = (-0.5, 0)$ does, indicating a different steady state.
    In this regime the linear theory $D_T^\infty = \text{Ca} + \mathcal{O} (\text{Ca}^2)$ breaks down completely.

\clearpage

\section{Droplets in Steady State}
\label{appendix:steady_state}
\subsection*{Weak to Moderate Shear Flows}
    \begin{figure}[h]
        \centering
        \hspace{0.05\linewidth}
        \begin{subfigure}{0.33\linewidth}
            \includegraphics[width=\linewidth]{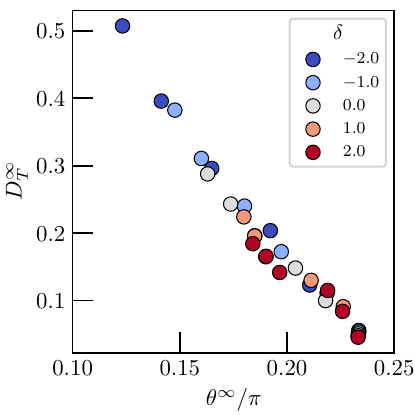}
            \caption{}
        \end{subfigure}
        \hspace{0.19\linewidth}
        \begin{subfigure}{0.33\linewidth}
            \centering
            \includegraphics[width=\linewidth]{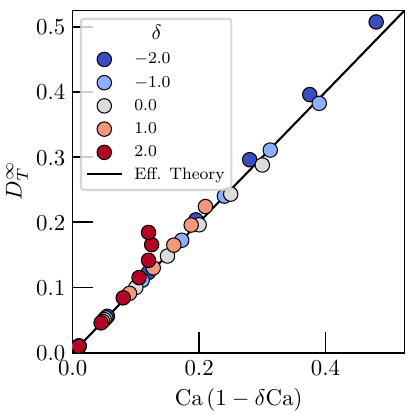}
            \caption{}
        \end{subfigure}
        \hspace{0.05\linewidth}
        \caption{
            (a) The steady state deformation parameter $D_T^\infty$ vs. the steady state orientation angle $\theta^\infty$ of the long axis of the droplet for different values of the odd viscosity difference $\delta$.
            (b) The steady state deformation parameter $D_T^\infty$ vs. the dimensionless group $\text{Ca}(1 - \delta \text{Ca})$.
            The effective theory $D_T^\infty = \text{Ca}(1 - \delta \text{Ca})$ is plotted in black.
            The parameters are $\text{Re} = 0.01$, $\lambda = 1$ and $L/R = 20$.
        }
        \label{fig:deform_vs_angle}
    \end{figure}
    Fig.~\ref{fig:deform_vs_angle}(a) shows the steady state deformation parameter $D_T^\infty$ as a function of the steady state orientation angle $\theta^\infty$ of the long axis of the droplet for different values of the odd viscosity difference $\delta$.
    There is a strong trend that shows no clear dependence on $\delta$.
    This is strong evidence that the orientation of the droplet cannot be influenced by odd viscosity independent of the degree of deformation.
    Thus simultaneous knowledge of $D_T^\infty$ and $\theta^\infty$ is likely insufficient to determine whether a material is odd.
    Fig.~\ref{fig:deform_vs_angle}(b) shows a purely phenomenological collapse of the data in Fig.~\ref{fig:deformation_theory}.
    The collapse suggests that $D_T^\infty \approx \text{Ca}(1 - \delta \text{Ca})$ for small $\delta$ and small $\text{Ca}$.
    A rigorous higher-order analytical solution is necessary to determine whether this result is the correct second order deformation theory.

\subsection*{Strong Shear Flows}
    \begin{figure}[h]
        \centering
        \includegraphics[width=\linewidth]{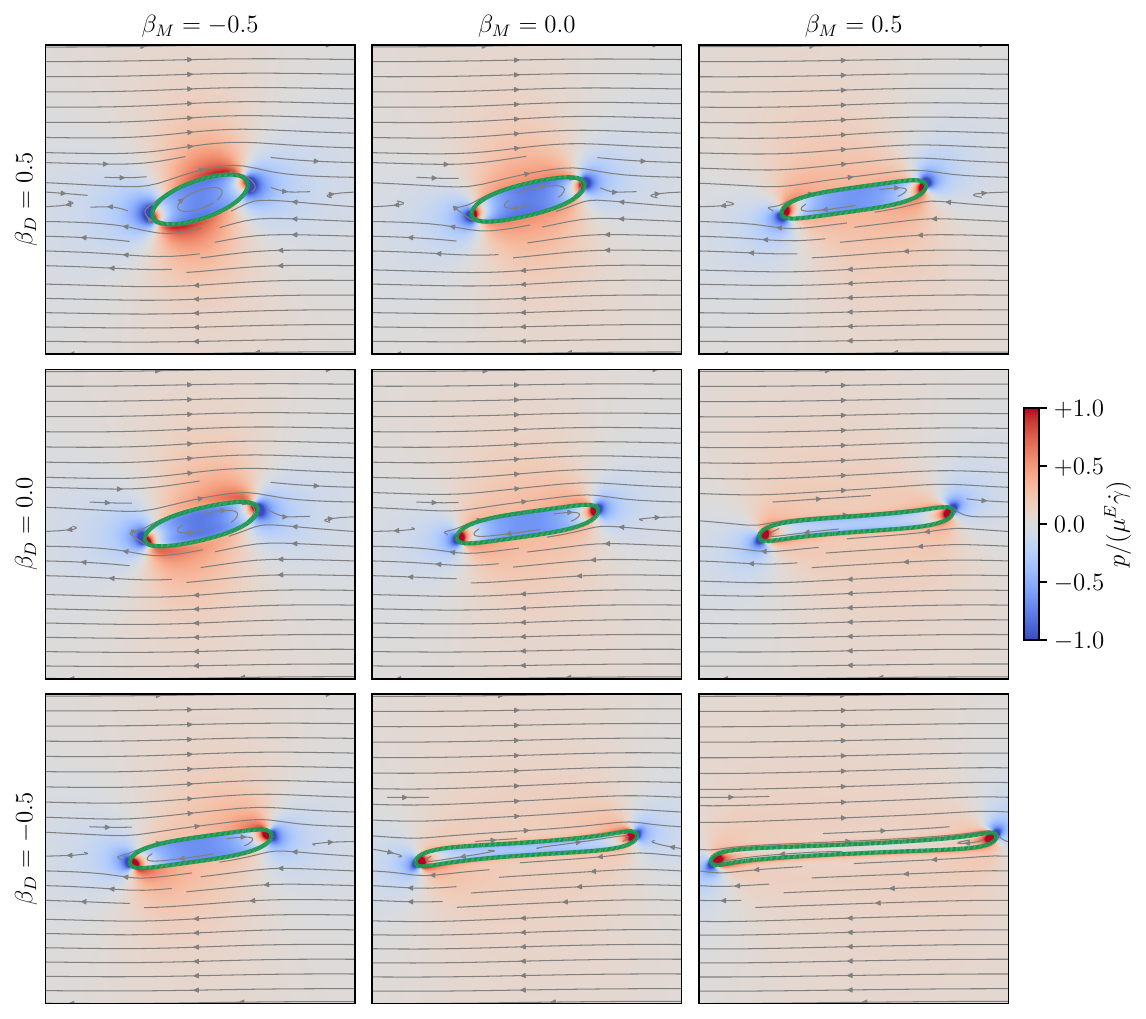}
        \caption{
            The pressure field $p$ for an odd-viscous droplet in an odd-viscous simple shear flow according to direct numerical simulations.
            The Laplace pressure, $\sigma/R$, has been subtracted from the droplet pressure field.
            The parameters are $\text{Re} = 0.01$, $\text{Ca} = 1.0$, $\lambda = 1$ and $L/R = 20$.
            Panels show different combinations of the oddness parameters $(\beta_D, \beta_M) \in [-0.5, 0.5] \times [-0.5, 0.5]$.
            The plotted region is $[-5R, 5R] \times [-5R, 5R]$.
            The grey curves denote streamlines of the velocity field $\bm{u}$.
        }
        \label{fig:deform_pressure_Ca=1.0}
    \end{figure}

    Fig.~\ref{fig:deform_pressure_Ca=1.0} shows the steady-state numerical solutions for strong shear ($\text{Ca} = 1.0$) and a range of dimensionless odd viscosities $-0.5 < \beta_D < 0.5$ and $-0.5 < \beta_M < 0.5$.
    The droplet shapes here depend much more strongly on the odd viscosities than in the moderate shear rate case ($\text{Ca} = 0.3$).
    That the flow and shape depend on $\delta$ and not $\beta_D$ or $\beta_M$ individually appears to break down, as is apparent when comparing panels $(\beta_D, \beta_M) = (-0.5, 0)$ and $(\beta_D, \beta_M) = (0, 0.5)$.
    Both cases have $\delta = -0.5$, yet the droplet with $(\beta_D, \beta_M) = (-0.5, 0)$ clearly has a larger value of $D_T^\infty$ than $(\beta_D, \beta_M) = (0, 0.5)$.
    The authors suspect this a numerical artifact due to the highly deformed interface, however further investigation is needed to rule out an emergent physics which occurs at large $\text{Ca}$.

\clearpage
    
\section{Validation}
\label{sec:validation}
    \begin{figure}[h]
        \centering
        \begin{subfigure}{\textwidth}
            \includegraphics[width=0.9\linewidth]{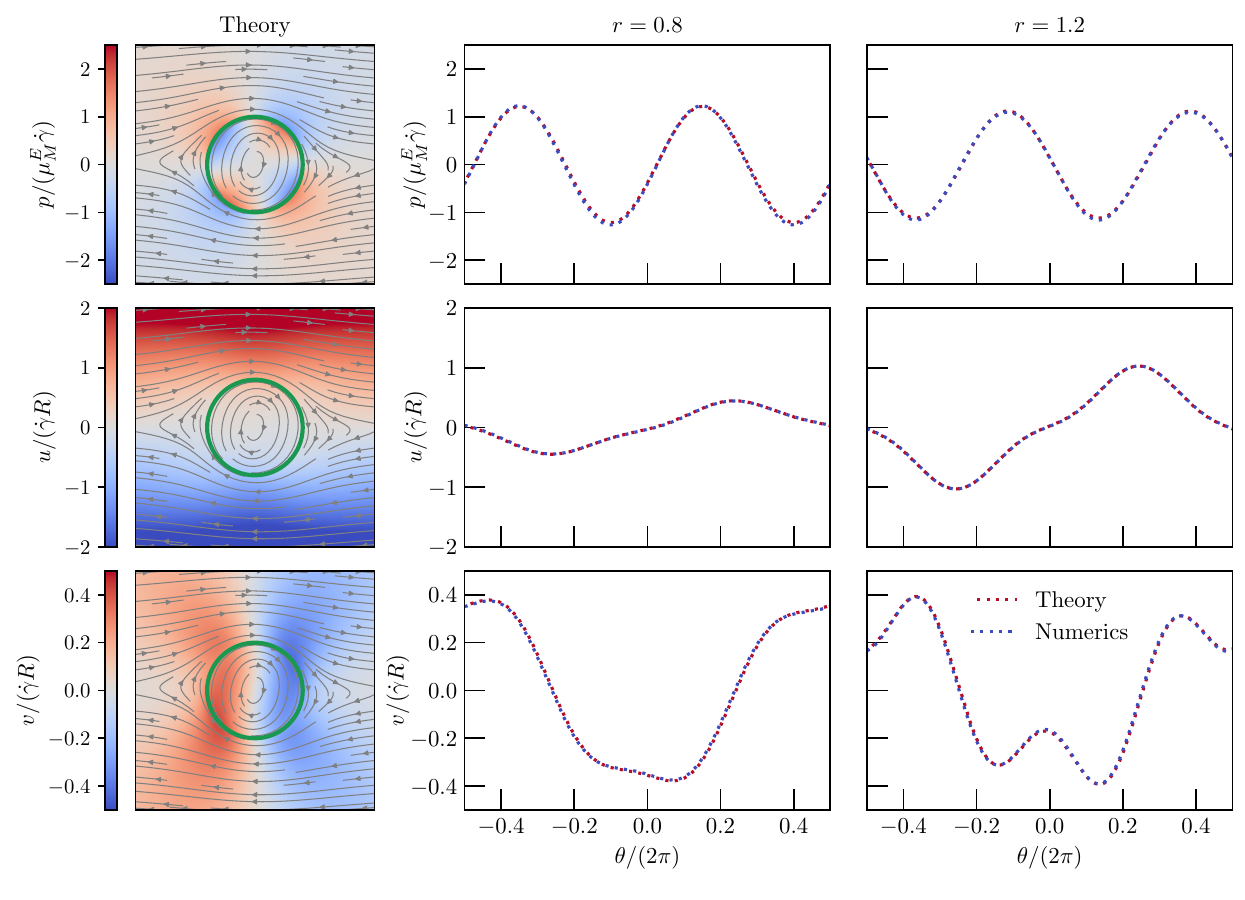}
            \subcaption{}
        \end{subfigure}
        \begin{subfigure}{\textwidth}
            \includegraphics[width=0.9\linewidth]{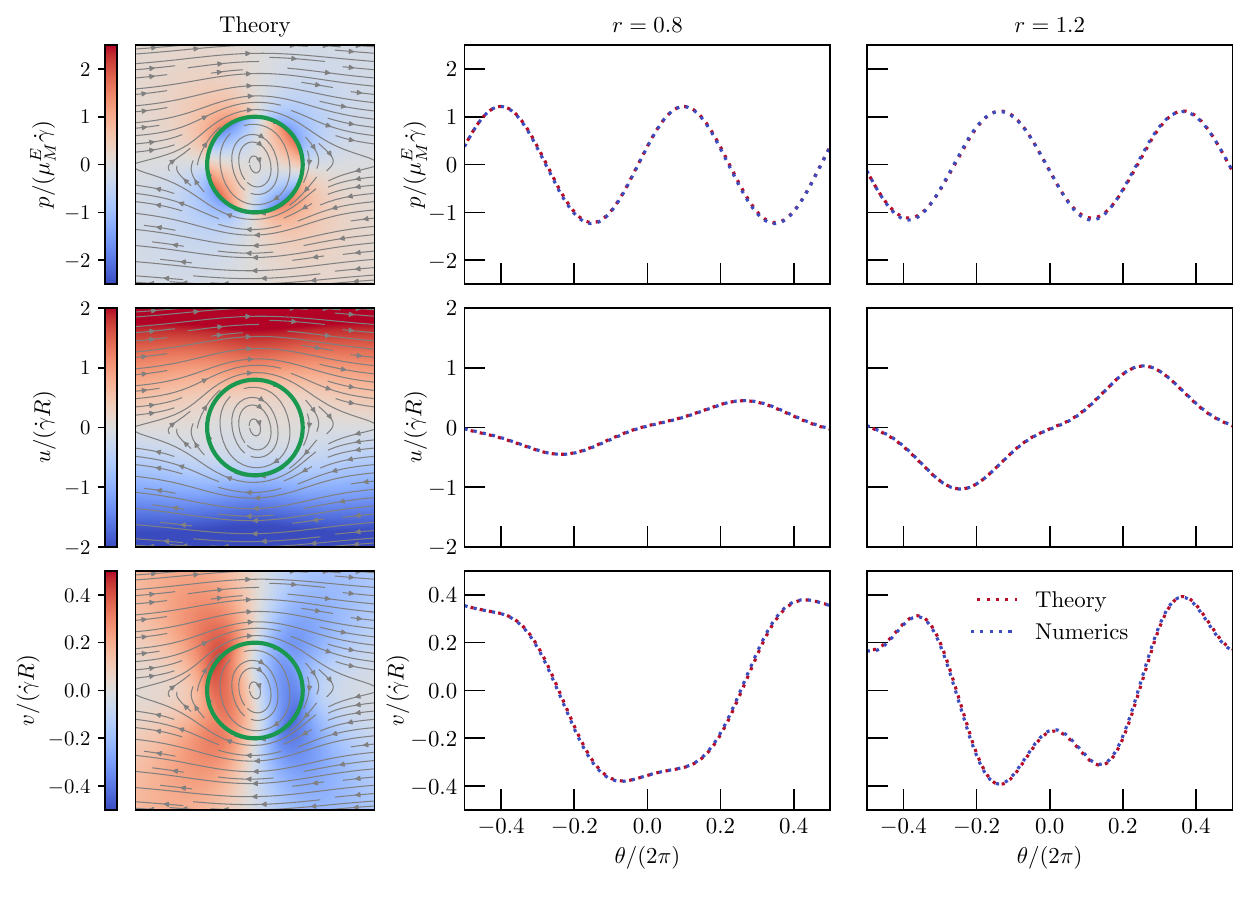}
            \subcaption{}
        \end{subfigure}
        \caption{
            Comparison between numerical and analytical solutions for the flow around an odd-viscous droplet.
            Shown are the Cartesian velocity components $u$ and $v$, and the pressure $p$.
            The Laplace pressure, $\sigma/R$, has been subtracted from the droplet pressure field.
            The first column displays $u$, $v$, and $p$ with streamlines for the analytical solution Eqs.~\eqref{eqn:u_r_int}-\eqref{eqn:p_ext}.
            The second and third columns show $u$, $v$, and $p$ sampled along circles of radius $r = 0.8$ and $r = 1.2$, respectively, centred at the origin, for both numerical and analytical solutions.
            For the numerical simulations, $\text{Re} = 0.01$, $\text{Ca} = 0.01$, $L/R = 20$ and $\texttt{LEVEL} = 11$.
            For the analytical solutions, $\text{Re} = 0$, $\text{Ca} = 0$, and the domain is taken to be infinite.
            Panels correspond to the odd-viscosity parameters:
            (a) $\beta_M = 0.0$, $\beta_D = 1.0$;
            (b) $\beta_M = 0.0$, $\beta_D = -1.0$.
        }
        \label{fig:validation}
    \end{figure}
    Fig.~\ref{fig:validation} shows a detailed comparison between the steady-state numerical solution and the odd Stokes flow solution Eqs.~\eqref{eqn:u_r_int}-\eqref{eqn:p_ext}. 
    The simulations have been validated for a variety $\beta_D$ and $\beta_M$ values between 0.0 and 0.5.
    Shown in Fig.~\ref{fig:validation} are
    $\beta_M$ =  0.0, $\beta_D$ =  1.0
    and
    $\beta_M$ =  0.5, $\beta_D$ =  -1.0.
    Included in the figure are both a visual comparison of the Cartesian components, $u$ and $v$ of the velocity, and the pressure $p$, as well as their values on circles of radius $r = 0.8R$ and $r = 1.2R$ respectively centred on the origin.
    The values $r = 0.8R$ and $r = 1.2R$ are chosen to emphasise the regions where the disturbance fields are largest.
    The Laplace pressure $\sigma/R$ has been subtracted from the droplet pressure.

\clearpage
\printbibliography

@article{Popinet2003,
  doi = {10.1016/s0021-9991(03)00298-5},
  url = {https://doi.org/10.1016/s0021-9991(03)00298-5},
  year = {2003},
  month = sep,
  publisher = {Elsevier {BV}},
  volume = {190},
  number = {2},
  pages = {572--600},
  author = {St{\'{e}}phane Popinet},
  title = {Gerris: a tree-based adaptive solver for the incompressible Euler equations in complex geometries},
  journal = {Journal of Computational Physics}
}

@article{Popinet2009,
  doi = {10.1016/j.jcp.2009.04.042},
  url = {https://doi.org/10.1016/j.jcp.2009.04.042},
  year = {2009},
  month = sep,
  publisher = {Elsevier {BV}},
  volume = {228},
  number = {16},
  pages = {5838--5866},
  author = {St{\'{e}}phane Popinet},
  title = {An accurate adaptive solver for surface-tension-driven interfacial flows},
  journal = {Journal of Computational Physics}
}

@article{Popinet2015,
  doi = {10.1016/j.jcp.2015.09.009},
  url = {https://doi.org/10.1016/j.jcp.2015.09.009},
  year = {2015},
  month = dec,
  publisher = {Elsevier {BV}},
  volume = {302},
  pages = {336--358},
  author = {St{\'{e}}phane Popinet},
  title = {A quadtree-adaptive multigrid solver for the Serre{\textendash}Green{\textendash}Naghdi equations},
  journal = {Journal of Computational Physics}
}

@article{
    Rallison_1980, title={Note on the time-dependent deformation of a viscous drop which is almost spherical}, volume={98}, DOI={10.1017/S0022112080000316}, number={3}, journal={Journal of Fluid Mechanics}, author={Rallison, J. M.}, year={1980}, pages={625–633}}

@BOOK{Lamb,
  title     = "Cambridge mathematical library: Hydrodynamics",
  author    = "Lamb, Horace",
  publisher = "Cambridge University Press",
  edition   =  6,
  month     =  nov,
  year      =  1993,
  address   = "Cambridge, England",
  language  = "en"
}

@article{Taylor1932,
    author = {Taylor, Geoffrey Ingram },
    title = {The viscosity of a fluid containing small drops of another fluid},
    journal = {Proceedings of the Royal Society of London. Series A, Containing Papers of a Mathematical and Physical Character},
    volume = {138},
    number = {834},
    pages = {41-48},
    year = {1932},
    doi = {10.1098/rspa.1932.0169},
    
    URL = {https://royalsocietypublishing.org/doi/abs/10.1098/rspa.1932.0169},
    eprint = {https://royalsocietypublishing.org/doi/pdf/10.1098/rspa.1932.0169}
}

@article{Taylor1934,
    author = {Taylor, Geoffrey Ingram },
    title = {The formation of emulsions in definable fields of flow},
    journal = {Proceedings of the Royal Society of London. Series A, Containing Papers of a Mathematical and Physical Character},
    volume = {146},
    number = {858},
    pages = {501-523},
    year = {1934},
    doi = {10.1098/rspa.1934.0169},
    
    URL = {https://royalsocietypublishing.org/doi/abs/10.1098/rspa.1934.0169},
    eprint = {https://royalsocietypublishing.org/doi/pdf/10.1098/rspa.1934.0169}
}

@article{BarthsBiesel1973,
  doi = {10.1017/s0022112073000534},
  url = {https://doi.org/10.1017/s0022112073000534},
  year = {1973},
  month = oct,
  publisher = {Cambridge University Press ({CUP})},
  volume = {61},
  number = {1},
  pages = {1--22},
  author = {D. Barth{\`{e}}s-Biesel and A. Acrivos},
  title = {Deformation and burst of a liquid droplet freely suspended in a linear shear field},
  journal = {Journal of Fluid Mechanics}
}

@article{rallison1984deformation,
   author = "Rallison, J M",
   title = "The Deformation of Small Viscous Drops and Bubbles in
Shear Flows", 
   journal= "Annual Review of Fluid Mechanics",
   year = "1984",
   volume = "16",
   number = "Volume 16, 1984",
   pages = "45-66",
   doi = "https://doi.org/10.1146/annurev.fl.16.010184.000401",
   url = "https://www.annualreviews.org/content/journals/10.1146/annurev.fl.16.010184.000401",
   publisher = "Annual Reviews",
   issn = "1545-4479",
   type = "Journal Article",
  }

@article{stone1994dynamics,
   author = "Stone, H A",
   title = "Dynamics of Drop Deformation and Breakup in Viscous
Fluids", 
   journal= "Annual Review of Fluid Mechanics",
   year = "1994",
   volume = "26",
   number = "Volume 26, 1994",
   pages = "65-102",
   doi = "https://doi.org/10.1146/annurev.fl.26.010194.000433",
   url = "https://www.annualreviews.org/content/journals/10.1146/annurev.fl.26.010194.000433",
   publisher = "Annual Reviews",
   issn = "1545-4479",
   type = "Journal Article",
  }

@article{Banerjee2017,
author = {Banerjee, Debarghya and Souslov, Anton and Abanov, Alexander G. and Vitelli, Vincenzo},
title = {Odd viscosity in chiral active fluids},
journal = {Nature Communications},
year = {2017},
volume = {8},
number = {1},
pages = {1573},
doi = {10.1038/s41467-017-01378-7},
url = {https://doi.org/10.1038/s41467-017-01378-7},
issn = {2041-1723},
note = {Published 2017-11-17}
}

@article{buckmaster1973bursting,
  title={The bursting of two-dimensional drops in slow viscous flow},
  author={Buckmaster, J. D. and Flaherty, J. E.},
  journal={J. Fluid Mech.},
  volume={60},
  number={4},
  pages={625--639},
  year={1973},
  DOI={10.1017/S0022112073000388},
  publisher={Cambridge University Press}
}

@article{acrivos1983breakup,
    author = {Acrivos, Andreas},
    title = {THE BREAKUP OF SMALL DROPS AND BUBBLES IN SHEAR FLOWS},
    journal = {Annals of the New York Academy of Sciences},
    volume = {404},
    number = {1},
    pages = {1-11},
    doi = {https://doi.org/10.1111/j.1749-6632.1983.tb19410.x},
    url = {https://nyaspubs.onlinelibrary.wiley.com/doi/abs/10.1111/j.1749-6632.1983.tb19410.x},
    eprint = {https://nyaspubs.onlinelibrary.wiley.com/doi/pdf/10.1111/j.1749-6632.1983.tb19410.x},
    year = {1983}
}

@article{richardson1968, 
    title={Two-dimensional bubbles in slow viscous flows},
    volume={33}, 
    DOI={10.1017/S0022112068001461}, number={3}, 
    journal={Journal of Fluid Mechanics}, author={Richardson, S.},
    year={1968}, 
    pages={475–493},
    publisher={Cambridge University Press}
}

@article{richardson1973two,
  title={Two-dimensional bubbles in slow viscous flows. Part 2},
  author={Richardson, S},
  journal={Journal of Fluid Mechanics},
  volume={58},
  number={1},
  pages={115--127},
  year={1973},
  DOI={10.1017/S0022112073002168},
  publisher={Cambridge University Press}
}

@article{Fruchart,
   author = "Fruchart, Michel and Scheibner, Colin and Vitelli, Vincenzo",
   title = "Odd Viscosity and Odd Elasticity", 
   journal= "Annual Review of Condensed Matter Physics",
   year = "2023",
   volume = "14",
   number = "Volume 14, 2023",
   pages = "471-510",
   doi = "https://doi.org/10.1146/annurev-conmatphys-040821-125506",
   url = "https://www.annualreviews.org/content/journals/10.1146/annurev-conmatphys-040821-125506",
   publisher = "Annual Reviews",
   issn = "1947-5462",
   type = "Journal Article",
  }

@Article{
    Avron1998,
    author={Avron, J. E.},
    title={Odd Viscosity},
    journal={Journal of Statistical Physics},
    year={1998},
    month={Aug},
    day={01},
    volume={92},
    number={3},
    pages={543-557},
    issn={1572-9613},
    doi={10.1023/A:1023084404080},
    url={https://doi.org/10.1023/A:1023084404080}
    }

@article{francca2025odd,
    title={Odd Droplets: Fluids with Odd Viscosity and Highly Deformable Interfaces}, 
    author={Hugo França and Maziyar Jalaal},
    year={2025},
    eprint={2503.21649},
    archivePrefix={arXiv},
    primaryClass={cond-mat.soft},
    journal={arXiv preprint},
    url={https://arxiv.org/abs/2503.21649}, 
}

@article{hosaka2021hydrodynamic,
  title = {Hydrodynamic lift of a two-dimensional liquid domain with odd viscosity},
  author = {Hosaka, Yuto and Komura, Shigeyuki and Andelman, David},
  journal = {Phys. Rev. E},
  volume = {104},
  issue = {6},
  pages = {064613},
  numpages = {10},
  year = {2021},
  month = {Dec},
  publisher = {American Physical Society},
  doi = {10.1103/PhysRevE.104.064613},
  url = {https://link.aps.org/doi/10.1103/PhysRevE.104.064613}
}

@article{Kogan,
  title = {Lift force due to odd Hall viscosity},
  author = {Kogan, E.},
  journal = {Phys. Rev. E},
  volume = {94},
  issue = {4},
  pages = {043111},
  numpages = {4},
  year = {2016},
  month = {Oct},
  publisher = {American Physical Society},
  doi = {10.1103/PhysRevE.94.043111},
  url = {https://link.aps.org/doi/10.1103/PhysRevE.94.043111}
}

@article{appleford2025rheology,
  title = {Rheology of two-dimensional dilute emulsions},
  author = {Appleford, Thomas and Sanjay, Vatsal and Jalaal, Maziyar},
  journal = {Phys. Rev. Fluids},
  volume = {11},
  issue = {3},
  pages = {033607},
  numpages = {19},
  year = {2026},
  month = {Mar},
  publisher = {American Physical Society},
  doi = {10.1103/9dgt-rkp5},
  url = {https://link.aps.org/doi/10.1103/9dgt-rkp5}
}

@article{duPlessis2025,
  author  = {du Plessis, Holly and Cosme, Pedro and 
             Fran{\c{c}}a, Hugo and Jalaal, Maziyar},
  title   = {Non-Reciprocal Capillary Waves},
  journal = {arXiv preprint},
  year    = {2025},
  eprint  = {2603.14195},
  archivePrefix = {arXiv},
  url     = {https://arxiv.org/abs/2603.14195}
}

@article{Khain_Scheibner_Fruchart_Vitelli_2022, title={Stokes flows in three-dimensional fluids with odd and parity-violating viscosities}, volume={934}, DOI={10.1017/jfm.2021.1079}, journal={Journal of Fluid Mechanics}, author={Khain, Tali and Scheibner, Colin and Fruchart, Michel and Vitelli, Vincenzo}, year={2022}, pages={A23}}

@article{Lier_2023,
  title = {Lift force in odd compressible fluids},
  author = {Lier, Ruben and Duclut, Charlie and Bo, Stefano and Armas, Jay and J\"ulicher, Frank and Sur\'owka, Piotr},
  journal = {Phys. Rev. E},
  volume = {108},
  issue = {2},
  pages = {L023101},
  numpages = {7},
  year = {2023},
  month = {Aug},
  publisher = {American Physical Society},
  doi = {10.1103/PhysRevE.108.L023101},
  url = {https://link.aps.org/doi/10.1103/PhysRevE.108.L023101}
}

@article{Lier_2024, title={Odd viscous flow past a sphere at low but non-zero Reynolds numbers}, volume={998}, DOI={10.1017/jfm.2024.915}, journal={Journal of Fluid Mechanics}, author={Lier, Ruben}, year={2024}, pages={A40}}

@article{Kirkinis,
  title = {Null-divergence nature of the odd viscous stress for an incompressible liquid},
  author = {Kirkinis, E.},
  journal = {Phys. Rev. Fluids},
  volume = {8},
  issue = {1},
  pages = {014104},
  numpages = {14},
  year = {2023},
  month = {Jan},
  publisher = {American Physical Society},
  doi = {10.1103/PhysRevFluids.8.014104},
  url = {https://link.aps.org/doi/10.1103/PhysRevFluids.8.014104}
}

@article{Yuan,
  title = {Stokesian dynamics with odd viscosity},
  author = {Yuan, Hang and Olvera de la Cruz, Monica},
  journal = {Phys. Rev. Fluids},
  volume = {8},
  issue = {5},
  pages = {054101},
  numpages = {29},
  year = {2023},
  month = {May},
  publisher = {American Physical Society},
  doi = {10.1103/PhysRevFluids.8.054101},
  url = {https://link.aps.org/doi/10.1103/PhysRevFluids.8.054101}
}

@article{Soni2019,
    author    = {Soni, Vishal and Bililign, Ephraim S. and Magkiriadou, Sofia and Sacanna, Stefano and Bartolo, Denis and Shelley, Michael J. and Irvine, William T. M.},
    title     = {The odd free surface flows of a colloidal chiral fluid},
    journal   = {Nature Physics},
    year      = {2019},
    volume    = {15},
    number    = {11},
    pages     = {1188--1194},
    month     = {11},
    doi       = {10.1038/s41567-019-0603-8},
    url       = {https://doi.org/10.1038/s41567-019-0603-8},
    issn      = {1745-2481}
}

@article{Lapa,
  title = {Swimming at low Reynolds number in fluids with odd, or Hall, viscosity},
  author = {Lapa, Matthew F. and Hughes, Taylor L.},
  journal = {Phys. Rev. E},
  volume = {89},
  issue = {4},
  pages = {043019},
  numpages = {13},
  year = {2014},
  month = {Apr},
  publisher = {American Physical Society},
  doi = {10.1103/PhysRevE.89.043019},
  url = {https://link.aps.org/doi/10.1103/PhysRevE.89.043019}
}

@article{hosaka2024chirotactic,
  title = {Chirotactic response of microswimmers in fluids with odd viscosity},
  author = {Hosaka, Yuto and Chatzittofi, Michalis and Golestanian, Ramin and Vilfan, Andrej},
  journal = {Phys. Rev. Res.},
  volume = {6},
  issue = {3},
  pages = {L032044},
  numpages = {8},
  year = {2024},
  month = {Aug},
  publisher = {American Physical Society},
  doi = {10.1103/PhysRevResearch.6.L032044},
  url = {https://link.aps.org/doi/10.1103/PhysRevResearch.6.L032044}
}
\end{document}